\documentclass[twocolumn,showpacs,preprintnumbers,amsmath,amssymb]{revtex4}
\usepackage{graphicx}% Include figure files
\usepackage{dcolumn}% Align table columns on decimal point
\usepackage{bm}% bold math
\begin{document}

%\preprint{APS/123-QED}

\title{Conserved Spin and Orbital Angular Momentum Hall Current in a Two-Dimensional Electron System with Rashba and Dresselhaus Spin-orbit Coupling}

\author{Tsung-Wei Chen,$^{1}$ Chih-Meng Huang,$^{1}$ and G.Y. Guo$^{1,2}$\footnote{Electronic address: gyguo@phys.ntu.edu.tw}}
\address{$^{1}$Department of Physics, National Taiwan University, Taipei 106, Taiwan\\
$^{2}$Department of Physics, Chinese University of Hong Kong, Shatin, N. T., Hong Kong}

\date{\today}% It is always \today, today,

\begin{abstract}
We study theoretically the spin and orbital angular momentum (OAM)
Hall effect in a high mobility two-dimensional electron system
with Rashba and Dresselhuas spin-orbit coupling by introducing
both the spin and OAM torque corrections, respectively, to the
spin and OAM currents. We find that when both bands are occupied,
the spin Hall conductivity is still a constant (i.e., independent
of the carrier density) which, however, has an opposite sign to
the previous value. The spin Hall conductivity in general
would not be cancelled by the OAM Hall conductivity. The OAM Hall
conductivity is also independent of the carrier density but
depends on the strength ratio of the Rashba to Dresselhaus
spin-orbit coupling, suggesting that one can manipulate the total Hall
current through tuning the Rashba coupling by a gate voltage. We
note that in a pure Rashba system, though the spin Hall
conductivity is exactly cancelled by the OAM Hall conductivity due
to the angular momentum conservation, the spin Hall effect could
still manifest itself as nonzero magnetization Hall current and
finite magnetization at the sample edges because the magnetic
dipole moment associated with the spin of an electron is twice as
large as that of the OAM. We also evaluate the electric field-induced
OAM and discuss the origin of the OAM Hall current. Finally, we
find that the spin and OAM Hall conductivities are closely related
to the Berry vector (or gauge) potential.
\end{abstract}

\maketitle
\section{Introduction}
Spin transport electronics or spintronics in semiconductors has
become a very active research field in condensed matter mainly
because of its potential applications in information storage and
processing and other electronic technologies~\cite{pri98} and also
because of many fundamental questions on the physics of electron
spin~\cite{zut04}. Spin current generation is an important issue
in the emerging spintronics. Recent proposals of the intrinsic
spin Hall effect are therefore remarkable~\cite{mur03,sin04}. In
the spin Hall effect, a transverse spin current is generated in
response to an electric field in a metal with relativistic
electron interaction (spin-orbit coupling). This effect has been
considered to arise extrinsically, i.e., by impurity
scattering~\cite{dya71}. The scattering becomes spin-dependent in
the presence of spin-orbit coupling, and this gives rise to the
spin Hall effect. In the recent proposals, in contrast, the spin
Hall effect can arise intrinsically in hole-doped ($p$-type) bulk
semiconductors~\cite{mur03} and also in electron-doped ($n$-type)
semiconductor heterostructures~\cite{sin04} due to intrinsic
spin-orbit coupling in the band structure. This intrinsic spin
Hall effect offers an exciting possibility of pure electric driven
spintronics in semiconductors, where spin-orbit coupling is
relatively strong and which can be more readily integrated with
well-developed semiconductor electronics.

A large number of theoretical papers have been written addressing
various issues about the intrinsic spin Hall effect. In
\cite{cul04}, a systematic semi-classical theory of spin transport
is presented, resolving a discrepancy between the prediction of
\cite{mur03} and the Kubo formula result.  In \cite{zha04}, an
orbital-angular-momentum (OAM) Hall current is predicted to exist
in response to an electric field and is found to cancel exactly
the spin Hall current in the spin Hall effect. In \cite{guo05},
however, {\it ab inito} relativistic band structure calculations
show that the OAM Hall conductivity in hole-doped semiconductors
is one order of magnitude smaller than the spin Hall conductivity,
indicating no cancellation between the spin and OAM Hall effects
in bulk semiconductors because of the orbital quenching by the
cubic crystalline anisotropy. There is also an intensive debate
about whether the intrinsic spin Hall effect remains valid beyond
the ballistic transport regime \cite{ino04,dim05,ocha05}. On the other hand,
experimental measurements of large spin Hall effects for the
Rashba two dimensional electron gas and for $n$-type bulk
semiconductors have just been reported~\cite{wun04,kat04},
although more work is needed to firmly establish the intrinsic or
extrinsic nature of the results.

At present, an urgent current issue in spintronics research is
about the appropriate definition of the spin
current~\cite{jin05,Sun05,zha05,wan05,wan05a}. In almost all
previous studies of the intrinsic spin Hall effect, the spin
current is intuitively defined as the expectation value of the
spin and velocity operators, namely, $\frac{1}{2}
\{\mathbf{v},s_z\}$, where $\{,\}$ is the anticommutator defined
by $\{A,B\}=AB+BA$. Similarly, the OAM current is defined as the
expectation value of $\frac{1}{2} \{\mathbf{v},L_z\}$. Here, $s_z$
and $L_z$ are the z-component of spin and OAM operators,
respectively, and $\mathbf{v}$ is the velocity operator. However,
this conventionally defined spin (OAM) current is not conserved in
systems with spin-orbit interaction~\cite{cul04}. Consequently,
many fundamental questions on the intrinsic spin Hall effect in
semiconductors remain unresolved. Very recently, in
Ref.\onlinecite{zha05}, a proper definition of the conserved
effective spin current is established for systems with spin-orbit
coupling, and the conserved spin current density is defined as
\begin{equation}
   \mathcal{J}_s=\mathbf{J}_s+\mathbf{P}_{\tau},
    \label{equ:current_def}
\end{equation}
where
$\mathbf{J}_s=Re[\Psi^{\dag}\frac{1}{2}\{\mathbf{v},s_z\}\Psi]$ is
conventional spin current density and $\mathbf{P}_{\tau}$ is the
torque dipole density which arise from spin torque
$\mathcal{T}^{s}_z=Re[\Psi^{\dag}\frac{1}{i\hbar}[s_z,H_0]\Psi]$
where $H_0$ is the Hamiltonian of the system. The commutator $[,]$
is defined by $[A,B]=AB-BA$. The effective spin current density
$\mathcal{J}_s$ then satisfies the standard continuity equation~\cite{zha05} of
$\frac{\partial \mathcal{S}_z}{\partial t}+\nabla\cdot
\mathcal{J}_s = 0$ where the spin density is defined by
$\mathcal{S}_{z}=\Psi^{\dag}s_z\Psi$. A derivation of the effective
spin continuity equation in Rashba-Dresselhaus system is given in
Appendix A. Within this definition of the effective spin current,
the unphysical intrinsic spin Hall effect in insulators with
localized orbitals vanishes, and the Onsager relation between the
spin Hall effect and inverse spin Hall effect is ensured.~\cite{zha05}
Furthermore, this new definition of the spin current predicts
opposite signs of spin Hall coefficients for a couple of
semiconductor models such as the Rashba and $k$-cubed Rashba
Hamiltonians.~\cite{zha05}

In this paper, we extend the theory proposed in Ref.
\onlinecite{zha05} to two-dimensional electron systems with both
Rashba and Dresselhaus spin-orbit coupling. Furthermore, we
introduce the concept of the effective OAM current by including an
OAM torque correction term, and investigate the OAM Hall effect
in both pure Rashba system and systems with Rashba-Dresselhaus
spin-orbit coupling. We derive the effective OAM continuity
equation for the Rashba-Dresselhaus system (Appendix B). Also
in this paper, we argue that in a pure Rashba system, though the
OAM Hall conductivity is found to exactly cancel the spin Hall
conductivity, there nevertheless would be nonzero magnetization
current and finite magnetization at the sample edges because the
magnetic dipole moment associated with spin angular moment is
twice as large as that of the OAM. Finally, we find that there are
interesting relations between the Berry vector potential and the
spin and OAM Hall conductivities.

\section{Model Hamiltonian and linear response calculation}
\subsection{Rashba-Dresselhaus Hamiltonian}
For a two-dimensional electron gas (2DEG) confined in a
semiconductor heterostructure, two major spin-orbit (SO)
interaction terms are usually present. One is the Rashba
term~\cite{byc84},
\begin{equation}
 H_R=\frac{\lambda}{\hbar}\vec{\sigma}\cdot(\mathbf{p}\times\hat{e}_{z}),
\end{equation}
which stems from the structural inversion asymmetry. Here,
$\vec{\sigma}=(\sigma _x,\sigma _y)$ and $\sigma _z$ are the three
Pauli matrices. The other is the Dresselhaus SO coupling which
results from the bulk inversion asymmetry, if the heterostructure
is made of semiconductors without spatial inversion symmetry such
as semiconductors in the zincblende structure~\cite{dre55}. The
Dresselhaus term is given by
\begin{equation}
H_D=-\frac{\beta}{\hbar}(p_x\sigma_x-p_y\sigma_y).
\end{equation}
Therefore, the full Hamiltonian for the 2DEG with Rashba-Dresselhaus SO coupling can be written as
\begin{equation}
     H_0=\frac{p^2}{2m}+H_R+H_D,
    \label{hamiltonian}
\end{equation}
where $m$ is the effective mass of the electrons in the 2DEG.
Interestingly, though the Dresselhaus coupling coefficient $\beta$
is fixed for a given structure, the Rashba coupling strength
$\lambda$ can be tuned by a gate voltage by up to 50
\%~\cite{nit97}, thereby providing an opportunity to study the interesting
interplay between both types of the SO coupling. The Hamiltonian
has been solved exactly by several authors. The eigenstates can be
written as
\begin{equation}
    |n\mathbf{k}\rangle=\frac{1}{\sqrt{2}}\begin{pmatrix}
        e^{-i\theta(\mathbf{k})} \\ in
    \end{pmatrix},
    \label{eigenstate_RD}
\end{equation}
where $n=\pm 1$ is band index and
\begin{equation}
    \tan\theta(\mathbf{k})=\frac{\lambda k_y-\beta k_x}{\lambda k_x-\beta k_y}.
    \label{tan_theta}
\end{equation}
 The corresponding eigenenergies is given by
\begin{equation}
    E_{n\mathbf{k}}=\frac{\hbar^2}{2m}k^2-nk \gamma(\phi),
    \label{eigenenergies_RD}
\end{equation}
where $\mathbf{k}=(k_x,k_y)$, $k=|\mathbf{k}|$,
$\tan(\phi)=k_y/k_x$, $\cos\theta(\mathbf{k})=(\lambda
\cos\phi-\beta \sin\phi)/\gamma(\phi)$,
$\sin\theta(\mathbf{k})=(\lambda \sin\phi-\beta
\cos\phi)/\gamma(\phi)$ and
$\gamma(\phi)=\sqrt{(\lambda^2+\beta^2)-2\lambda\beta
\sin(2\phi)}$. It can be proved that the difference of two Fermi
momenta $\hbar k^{\pm}_{F}$ is given by
($k^{+}_{F}-k^{-}_{F})=2m\gamma(\phi)/\hbar^2$. The band structure
consists of two energy bands which are degenerate at the centre of
the 2D momentum space ${\bf k} = 0$, as schematically illustrated
in the left panel of Fig. 1(a). The spins associated with the
eigenstates all lie in the $xy$ plane, as shown in the right panel
of Fig. 1(a) and in the left panel of Fig. 1(b).

\begin{figure}
\begin{center}
\includegraphics{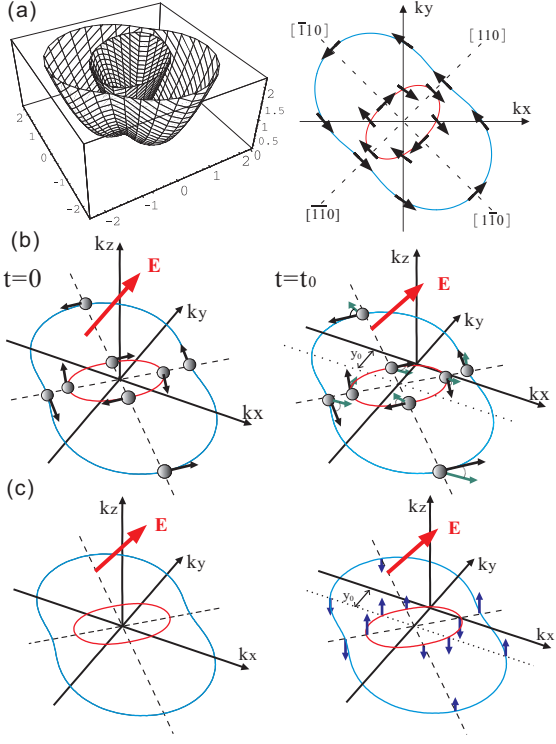}
\end{center}
\caption{(Color online) (a) The schematic bandstructure of the
Rashba-Dresselhaus spin-orbit coupling system with
$|\lambda|\neq|\beta|$. The outer sheet is for $n = +1$ and the
inner sheet, for $n=-1$. The two sheets touch at ${\bf k}=0$. If
$\lambda = \beta$, the sheets become degenerate at line $k_x =
k_y$ (see, also, Fig. 1 in~\cite{she04}). The right panel is the
top view of the two bands. The arrows labeled on the two bands
denote the spin directions of the associated $k$-points,
respectively. (b) Initially, the spins are in the $xy$-plane (the
left panel). At a short time $t_0$ after an external electric field $E$
(indicated as the large horizontal (red) arrow) is applied along the $y$ axis,
the bands move in the -$k_y$ direction with the
distance $\delta k_y=eEt_0/\hbar$ (see Subsec. II.E).
Each spinor feels an effective
magnetic field and will do precession. For $k_x>0$, the spins tend
to tilt down and for $k_x<0$, the spins tend to tilt up (the right
panel). The $z$ component of the spin is given by $\delta \langle
s_z \rangle =-eE\hbar (\lambda ^2-\beta ^2) \cos \phi /4nk^2
\gamma^3(\phi)$ (c) Before an external electric field is applied, all the
eigenstates carry zero orbital angular momentum (OAM), i.e.,
$\langle L_z \rangle _0=0$ (see Subsec. II.E.). When the electric
field is turned on, the $z$ component of the OAM is induced as shown
in the right panel (small vertical arrows).
A linear response calculation would give the
$z$ component as: $\delta \langle L_z \rangle = eE\hbar (\lambda
^2-\beta ^2)^2 \cos \phi /4nk^2 \gamma^5(\phi)$.}
\end{figure}

\subsection{Spin Hall conductivity}
Let us consider an uniform electric field $\mathbf{E}$ applied in
the $y$-direction. The total Hamiltonian in this case is given by
$H=H_0+eEy$, where $-e$ is the electron charge. Let us treat the
term $eEy$ as a small perturbation. To determine the spin
transport coefficient, we start with the linear response Kubo
formula in the clean limit. The conventional spin Hall
conductivity is~\cite{mar00}
\begin{equation}
    \begin{split}
    \sigma^{s_0}_{xy}&=-\frac{e\hbar}{V}\sum_{n\neq n'}\sum_{\mathbf{k}}[f_{n \mathbf{k}}-f_{n'\mathbf{k}}]\\
    &\times \frac{Im[\langle n\mathbf{k}|j^{s_z}_{x}|n'\mathbf{k}\rangle\langle n'\mathbf{k}|v_{y}|n\mathbf{k}\rangle]}{(E_{n\mathbf{k}}-E_{n'\mathbf{k}})^2},
        \end{split}
    \label{equ:kubo_1}
\end{equation}
where $j^{s_z}_{x}=1/2\{v_x,s_z\}$ is the usual spin current
operator, $\mathbf{v}(\mathbf{k})=\partial
H_0(\mathbf{k})/\hbar\partial \mathbf{k}$,
$H_0(\mathbf{k})=e^{-i\mathbf{k}\cdot
\mathbf{x}}H_0e^{i\mathbf{k}\cdot \mathbf{x}}$,
$s_z=\frac{\hbar}{2}\sigma_z$, $|n\mathbf{k}\rangle$ is the
eigenstate of the $n$th band with momentum $\mathbf{k}$, and
$f_{n\mathbf{k}}$ is the Fermi-Dirac distribution. The index
$n\neq n'$ denotes no intraband transition. Similarly, the spin
torque response coefficient can also be obtained~\cite{zha05}:
\begin{equation}
    \begin{split}
    \mathbf{\chi}(\mathbf{q})&=\frac{ie\hbar}{V}\sum_{n \neq n'}\sum_{\mathbf{k}}[f_{n\mathbf{k}}-f_{n'\mathbf{k}+\mathbf{q}}]\\
                 &\times
                 \frac{\langle n\mathbf{k}|\tau_{s}(\mathbf{k},\mathbf{q})|n'\mathbf{k}+\mathbf{q}\rangle\langle n'\mathbf{k}+\mathbf{q}|\mathbf{v}(\mathbf{k},\mathbf{q})|n\mathbf{k}\rangle}{(E_{n\mathbf{k}}-E_{n'\mathbf{k}+\mathbf{q}})^2},
    \end{split}
    \label{equ:kubo_2}
\end{equation}
where
$\tau_{s}(\mathbf{k},\mathbf{q})=\frac{1}{2}[\tau_{s}(\mathbf{k})+\tau_{s}(\mathbf{k}+\mathbf{q})]$,
$\mathbf{v}(\mathbf{k},\mathbf{q})=\frac{1}{2}[\mathbf{v}(\mathbf{k})+\mathbf{v}(\mathbf{k}+\mathbf{q})]$
with $\tau_{s}=\frac{1}{i\hbar}[s_z,H_0]$. In the above equation,
$\sigma^{s_0}_{xy}$ is the transport coefficient of conventional
spin current and $\sigma^{\tau_{s}}_{xy}$ can be determined from
the $dc$ response of the spin torque dipole:
$\mathbf{P}_{\tau}=Re[i\nabla_{\mathbf{q}}\chi(\mathbf{q})]|_{\mathbf{q}\rightarrow
0}\cdot \mathbf{E}$, namely, $\sigma^{\tau_{s}}_{xy}=Re[i\partial
\chi_{y}(\mathbf{q})/\partial q_x]|_{q_x\rightarrow
0}$~\cite{zha05}. Here $\mathbf{\chi}(\mathbf{q})$ is the response
coefficient of spin torque to an electric field with finite
wavevector $\mathbf{q}$. Before doing summation in the above Kubo
formula, we expand the term inside the summation to the first
order of $q$. At zero temperature,
$f_{n'\mathbf{k}+\mathbf{q}}\simeq
f_{n'\mathbf{k}}+\mathbf{q}\cdot(\partial
E_{n'\mathbf{k}}/\partial\mathbf{k})
\partial f_{n'\mathbf{k}}/\partial E_{n'\mathbf{k}}$, where the first
term is a step function while the second term was involved with a
Dirac delta function. We assume that the two bands are all
occupied by the electrons and the Fermi energy is larger than the
SO splitting. The spin Hall conductivity is
\begin{equation}
    \sigma^{s}_{xy}=\sigma^{s_0}_{xy}+\sigma^{\tau_s}_{xy},
    \label{equ:trans_coeff}
\end{equation}
which is the linear response of the conserved effective spin
current $\mathcal{J}_s$ to the electric field~\cite{zha05}. It is
straightforward to calculate the spin Hall conductivity (the
detailed derivation is given in Appendix C).
We find that $\sigma^{s_0}_{xy}=-sign (\lambda ^2-\beta^2)e/8\pi$ is the
conventional spin Hall conductivity and $\sigma^{\tau_s}_{xy}=sign
(\lambda ^2-\beta^2)e/4\pi$ is the conductivity due to the spin
torque correction. Therefore, the spin Hall conductivity defined
as the response of the conserved effective spin current is:
\begin{equation}
    \sigma^{s}_{xy}=\left\{ \begin{array}{rcl}
    \frac{e}{8\pi}, & \mbox{} & \lambda^2>\beta^2 \\
        0,           & \mbox{} & \lambda^2=\beta^2      \\
    -\frac{e}{8\pi},  & \mbox{} & \lambda^2<\beta^2, \\
    \end{array}\right.
\end{equation}
which is independent on the carrier density. Remarkably, with the
spin torque correction term $\sigma^{\tau_s}_{xy}$ included, the
sign of the effective spin Hall conductivity is opposite to the
conventional spin Hall conductivity reported in \cite{she04} and
\cite{sin04a}.

\subsection{Orbital angular momentum Hall conductivity}
%The spin Hall effect is essentially caused by the SO coupling in the absence of magnetic field.
%How about the OAM Hall effect?
The usual OAM current operator was introduced as
\begin{equation}
    j^{o_0}_{x}=\frac{1}{2}\{v_x,L_z\},
    \label{equ:oam_op}
\end{equation}
where $L_z$ is the $z$ component of the OAM operator. In the
Rashba Hamiltonian and with this definition of the OAM current,
the spin Hall effect is always accompanied by the OAM Hall effect
which was first noted by Zhang and Yang~\cite{zha04}. In a SO
system, such a definition has the same problems as the
conventional spin current operator because the OAM current is not
a conserved quantity. Therefore, we derive the continuity equation
for the effective OAM current (see Appendix B). We
find $\frac{\partial\mathcal{L}_z}{\partial t}+\nabla\cdot
\mathbf{J}_o=\mathcal{T}^{o}_z$, where
$\mathcal{L}_z=\Psi^{\dag}L_z\Psi$ is the $z$ component of the OAM
density, $\mathbf{J}_o=Re[\Psi^{\dag}\mathbf{j}^{o_0}\Psi]$ is the OAM
current density and
$\mathbf{j}^{o_0}=\frac{1}{2}\{\textbf{v},L_{z}\}$.
$\mathcal{T}^{o}_z=Re[\Psi^{\dag}\tau_{o}\Psi]$ is the torque
density, where $\tau_{o}=\frac{1}{i\hbar}[L_{z},H_0]$. In analogue
to the conserved effective spin current operator proposed
in~\cite{zha05}, the average torque in the bulk is zero,  and hence we have
$\frac{1}{V}\int dV\mathcal{T}^{o}_z=0$. The torque density can be
written as the divergence of torque dipole density $\mathbf{P}^{o}_{\tau}(\mathbf{x})$, i.e.,
$\mathcal{T}^{o}_z=-\nabla\cdot\mathbf{P}^{o}_{\tau}(\mathbf{x})$.
%where $\mathbf{P}^{o}_{\tau}(\mathbf{x})$ is the torque dipole density.
Substituting
$\mathcal{T}^{o}_z=-\nabla\cdot\mathbf{P}^{o}_{\tau}$ into
continuity equation of OAM, we obtain
$\mathcal{J}_o=\mathbf{J}_o+\mathbf{P}^{o}_{\tau}$ as
the effective OAM current. The torque dipole density vanishes outside
the bulk, and we can write
$\int_{V}dV\mathbf{P}^{o}_{\tau}=\int_{V}dV(-\mathbf{x}\nabla\cdot\mathbf{P}^{o}_{\tau})=\int_{V}dV\mathbf{x}\mathcal{T}^{o}_z$.
Therefore, the unique form of torque dipole density is
$\mathbf{P}^{o}_{\tau}=Re[\Psi^{\dag}(\mathbf{x}\tau_o)\Psi]$. In short, we
may define an effective OAM current operator:
\begin{equation}
    \hat{\mathcal{J}}_o=\frac{d\mathbf{x}}{dt}L_z+\mathbf{x}\tau_o.
\end{equation}
It has an extra term $\mathbf{x}\tau_o$ which is the correction
term due to the OAM torque. The corresponding OAM current density
$\mathcal{J}_o=Re\Psi^{\dag}(\mathbf{x})\hat{\mathcal{J}}_o\Psi(\mathbf{x})$
defined as the expectation value of this current operator
satisfies the standard continuity equation of $ \frac{\partial
\mathcal{L}_z}{\partial t}+\nabla\cdot \mathcal{J}_o = 0$. As for
the spin Hall conductivity~\cite{zha05}, the OAM Hall conductivity
has two parts:
\begin{equation}
\sigma^{o}_{xy}=\sigma^{o_0}_{xy}+\sigma^{\tau _{o}}_{xy},
\end{equation}
where the first term is the usual OAM Hall conductivity and the
second term  comes from the OAM torque correction. Thus, the OAM
Hall conductivity can be calculated from Eq. (\ref{equ:kubo_1})
and Eq. (\ref{equ:kubo_2}) by substituting $j^{s_z}_{x}$ and
$\tau_{s}$ with $j^{o_0}_{x}$ and
$\tau_{o}=\frac{1}{i\hbar}[L_z,H_0]$, respectively. It can be
shown (Appendix C) that the OAM Hall conductivity defined as the
linear response of the effective OAM current is:
\begin{equation}
    \sigma^{o}_{xy}=\left\{ \begin{array}{rcl}
    -\frac{\lambda^2+\beta^2}{\lambda^2-\beta^2}\frac{e}{8\pi}, & \mbox{} & \lambda^2>\beta^2 \\
        0,               & \mbox{} & \lambda^2=\beta^2      \\
        \frac{\lambda^2+\beta^2}{\lambda^2-\beta^2}\frac{e}{8\pi},  & \mbox{} & \lambda^2<\beta^2 \\
    \end{array}\right .
\end{equation}
which is a function of the ratio $\lambda/\beta$ and is also
independent of the carrier density.

\subsection{Berry vector potential and Hall conductivity}

When the Hamiltonian of a physical system is parameterized by
periodically changing environment, the state ket of system will
travel on a close path and return to initial state ket after a
period. The final state ket must coincide with initial state
vector, apart from a phase factor. Berry~\cite{ber84} has shown
that the state ket will acquire an additional phase factor as the
system undergoes the adiabatic evolution. The phase factor accompanying
the adiabatic evolution is called Berry phase. Berry phase cannot be
removed by any gauge transformation for a closed
path~\cite{boh03}. In solids, the Bloch state also acquire a Berry
phase if the applied perturbation can make a constraint such that
$\mathbf{k}$ travels adiabatically on a closed path in the
Brillouin zone~\cite{Zak89}. The Berry phase of Bloch states
$|n\mathbf{k}\rangle$ for a closed path $C$ can be expressed by
\begin{equation}
\Phi_{n}=\oint_{c}\mathbf{A}_{n}(\mathbf{k})\cdot d\mathbf{l},
\label{Berryphase}
\end{equation}
where $\mathbf{A}_{n}(\mathbf{k})=\langle
n\mathbf{k}|(-i)\frac{\partial}{\partial\mathbf{k}}|n\mathbf{k}\rangle$
is defined as Berry vector potential (or Berry connection).
%$n$ is the band index.
Berry curvature is defined as
$\Omega_{n}=\nabla_{\mathbf{k}}\times\mathbf{A}_n$. The Berry
vector potential and curvature are the salient characteristic of
energy band structure and hence have important applications in
transport properties of carriers. It has been shown that the equation of
motion of Bloch electron has an extra anomalous velocity in terms
of Berry curvature of Bloch states~\cite{Cha95}. In bulk $p$-type
semiconductors with spin-orbit coupling, the existence of $k$-space
magnetic monopole (Berry curvature) in the degeneracy $\Gamma$
point of band structure results in a transverse force exerting on
spin~\cite{mur03}. The connection between the dissipationless spin
Hall conductivity and Berry curvature has been shown
in~\cite{Mur04}. Here we further show that in
two-dimensional Rashba-Dresselhaus systems, we can write the spin Hall and OAM
Hall conductivities in terms of Berry curvature and vector potential.

%In Ref. \cite{she04}, the spin Hall conductivity in
%the Rashba-Dresselhaus system is found to be related to the Berry
%phase. Here we further show that the spin and OAM Hall
%conductivities are closely related to the Berry vector potential
%(or Berry connection).

With the eigenstates $|n\mathbf{k}\rangle$ given by Eq.
(\ref{eigenstate_RD}), we can prove that the matrix element
$\langle n'\mathbf{k}|v_y|n\mathbf{k}\rangle$ satisfies the
following relation (for $n'\neq n$):
\begin{equation}
\langle
n'\mathbf{k}|v_y|n\mathbf{k}\rangle=\frac{i}{\hbar}(E_{n\mathbf{k}}-E_{n'\mathbf{k}})\langle
n'\mathbf{k}|(-i)\frac{\partial}{\partial k_y}|n\mathbf{k}\rangle
\label{yvelocity}
\end{equation}
where $\langle n'\mathbf{k}|(-i)\frac{\partial}{\partial
k_y}|n\mathbf{k}\rangle=-\frac{1}{2}\frac{\partial\theta}{\partial
k_y}$.
The Berry vector potential is then given by
\begin{equation}
\mathbf{A}=\langle
n\mathbf{k}|(-i)\frac{\partial}{\partial\mathbf{k}}|n\mathbf{k}\rangle=-\frac{1}{2}\frac{\partial\theta}{\partial\mathbf{k}}.
\label{partialk}
\end{equation}
Substituting Eqs. (\ref{yvelocity}), and (\ref{partialk})
into Eq. (\ref{equ:kubo_1}) and noting that
$\sigma^{s_0}_{xy}=-\sigma^{s_0}_{yx}$~\cite{cond}, we have
\begin{equation}
\sigma^{s_0}_{xy}=-\frac{e\hbar}{2mV}\sum_{n}\sum_{\mathbf{k}}\frac{f_{n\mathbf{k}}}{\omega_{n\mathbf{k}}}[\mathbf{k}\times\mathbf{A}]_{z},
\label{berrspin}
\end{equation}
where $\omega_{n\textbf{k}}$ is defined as
$\omega_{n\textbf{k}}\equiv(E_{n\textbf{k}}-E_{-n\textbf{k}})/\hbar$.
Similarly, we also obtain,
\begin{equation}
  \sigma^{o_0}_{xy}=-\frac{e\hbar}{mV}\sum_{n}\sum_{\mathbf{k}}\frac{f_{n\mathbf{k}}}{\omega_{n\mathbf{k}}}[\mathbf{k}\times\mathbf{A}]_{z}^2,
\label{berroam}
\end{equation}
where $\sigma^{o_0}_{xy}=-\sigma^{o_0}_{yx}$ was used~\cite{cond}.
The spin Hall and OAM Hall conductivities are therefore related to Berry
vector potential. The Berry curvature is
$\Omega(\textbf{k})_{z}
=[\nabla_{\mathbf{k}}\times\mathbf{A}]_z=-\frac{(\lambda^2-\beta^2)}{|\lambda^2-\beta^2|}\pi\delta^{(2)}(\textbf{k})$.
The corresponding Berry phase is then given by the path integral of the Berry
vector potential, i.e., $\Phi=\oint_{C}
d\textbf{l}\cdot\mathbf{A}=\int_{R}\Omega(\textbf{k})_{z}dk_xdk_y
=-\frac{\lambda^2-\beta^2}{|\lambda^2-\beta^2|}\pi=-sign(\lambda^2-\beta^2)\pi$,
where $R$ is the region containing the origin in the {\bf
k}-plane. The Berry curvature $\Omega_{z}$ is zero for
$\textbf{k}\neq 0$ but non-zero for $\textbf{k}= 0$ where the two
bands of the Rashba-Dresselhaus system are degenerate.
As for the spin transverse force for the Luttinger Hamiltonian given in
Ref. \onlinecite{Mur04}, the corresponding spin transverse force in
Rashba-Dresselhaus system can also be given in terms of the Berry curvature.
Even though the Berry curvature vanishes for
$\mathbf{k}\neq 0$, the spin Hall current can nonetheless occur due to
the Aharonov-Bohm-like effect at $\mathbf{k}=0$~\cite{Sch05}.

%In the pure Rashba system, the Berry vector potential is $A_y(k)=-\cos\phi/(2k)$,
%$A_x(k)=\sin\phi/(2k)$, and the energy difference is
%$\omega_{n\textbf{k}}=-2n\lambda k/\hbar$. Consequently,
%$[\mathbf{k}\times\mathbf{A}(\textbf{k})]_{z}$ is a constant in
%pure Rashba system:
%\begin{equation}
%[\mathbf{k}\times\mathbf{A}(\textbf{k})]_{z}=-\frac{1}{2}.
%\label{Ra}
%\end{equation}
%Assuming the Fermi level is larger than the spin-orbit coupling,
%we get an universal value of $-\frac{e}{8\pi}$ for the spin Hall
%conductivity. For the Rashba-Dresselhaus system, we have:
% \begin{equation}
% [\mathbf{k}\times\mathbf{A}(\textbf{k})]_{z}=-\frac{1}{2}\frac{(\lambda^2-\beta^2)}{[\gamma(\phi)]^2}.
% \end{equation}
%From Eqs. (\ref{berrspin}), (\ref{berroam}) and (\ref{Ra}), we can
%arrive at $(\sigma^{s_0}_{xy}+\sigma^{o_0}_{xy})=0$ for the Rashba
%system~\cite{zha04}. For the Rashba-Dresselhaus system, the
%formula (16) gives the same conventional spin Hall conductivity of
%$-sign(\lambda^2-\beta^2)e/8\pi$, as has been reported
%in~\cite{she04,sin04a}.

Let us now discuss the effects of the different choice of the eigenstates on
the above findings. In, e.g., Ref. ~\cite{Cha05}, the eigenstates
($|\widetilde{n\mathbf{k}}\rangle$)
\begin{equation}
|\widetilde{+\mathbf{k}}\rangle=\frac{1}{\sqrt{2}}\left(\begin{array}{c}
-ie^{-i\theta}\\
1
\end{array}\right);
|\widetilde{-\mathbf{k}}\rangle=\frac{1}{\sqrt{2}}\left(\begin{array}{c}
1\\
-ie^{i\theta}
\end{array}\right).
\label{eigenstate RD2}
\end{equation}
were used for the Rashba-Dresselhaus system.
We find that the unitary transformation defined by
the matrix
\begin{equation}
U=\frac{1}{2}\left(\begin{array}{cc}
-i+e^{i\theta}&i-e^{-i\theta}\\
e^{i\theta}-ie^{2i\theta}&-i+e^{i\theta}\\
\end{array}\right)
\label{UT}
\end{equation}
with $UU^{\dag}=U^{\dag}U=1$, will transform the eigenstates of
Eq. (\ref{eigenstate_RD}) to that of Eq. (\ref{eigenstate RD2}),
i.e., $|\widetilde{n\mathbf{k}}\rangle=U|n\mathbf{k}\rangle$.
%Applying the
%unitary matrix $U$ to the eigenstates in Eq.
%(\ref{eigenstate_RD}), we obtain the other eigenstate denoted as
%$|\widetilde{n\mathbf{k}}\rangle=U|n\mathbf{k}\rangle$ :
%\begin{equation}
%|\widetilde{+\mathbf{k}}\rangle=\frac{1}{\sqrt{2}}\left(\begin{array}{c}
%-ie^{-i\theta}\\
%1
%\end{array}\right);
%|\widetilde{-\mathbf{k}}\rangle=\frac{1}{\sqrt{2}}\left(\begin{array}{c}
%1\\
%-ie^{i\theta}
%\end{array}\right).
%\label{eigenstate RD2}
%\end{equation}
Using the eigenstates in Eq. (\ref{eigenstate RD2}), the Berry
vector potential is given by
\begin{equation}
\begin{split}
\mathbf{A}_{+}\equiv\langle
\widetilde{+\mathbf{k}}|(-i)\frac{\partial}{\partial\mathbf{k}}|\widetilde{+\mathbf{k}}\rangle&=-\frac{1}{2}\frac{\partial\theta}{\partial\mathbf{k}}\\
\mathbf{A}_{-}\equiv\langle
\widetilde{-\mathbf{k}}|(-i)\frac{\partial}{\partial\mathbf{k}}|\widetilde{-\mathbf{k}}\rangle&=+\frac{1}{2}\frac{\partial\theta}{\partial\mathbf{k}}
\end{split}
\end{equation}
that depends on the band index $n$.  We may define
$\mathbf{A}_n=n\mathbf{A}$ for both the bases of (\ref{eigenstate_RD}) and
(\ref{eigenstate RD2}). As a result, the Berry phase given by Eq.
(\ref{Berryphase}) also depends on the band index $n$ \cite{Cha05}, i.e.,
\begin{equation}
\Phi_{\pm}=\oint_{c}\mathbf{A}_{\pm}(\mathbf{k})\cdot
d\mathbf{l}=\mp sign(\lambda^2-\beta^2)\pi=\pm\Phi,
\end{equation}
where $\Phi$ is the Berry phase when the basis of Eq.
(\ref{eigenstate_RD}) is used. In contrast, the signs of the Berry phase $\Phi$ for
the two bands are the same in the choice of the eigenstates of Eq.
(\ref{eigenstate_RD}). Nevertheless, we can write $\Phi_n=n\Phi$ with $n=\pm
1$. Note that the charge Hall effect would be zero when the basis
of Eq. (\ref{eigenstate RD2}) is used, and would be nonzero if the
basis of Eq. (\ref{eigenstate_RD}) is chosen, as have been pointed out
in Ref. \onlinecite{Cha05}.

%Using eigenstates in Eq.
%(\ref{eigenstate RD2}), we can evaluate the spin Hall conductivity
%and OAM Hall conductivity based on Eq. (\ref{equ:kubo_1}) and the
%results are in agreement with Eq. (\ref{berrspin}) and Eq.
%(\ref{berroam}) respectively.

Using
$\mathbf{A}=-\frac{1}{2}\frac{\partial\theta}{\partial\mathbf{k}}$
(Eq. \ref{partialk}), we can show that $\textbf{A}=[\textbf{k}\times
\textbf{A}]_{z}\textbf{a}_{0}$, where
$\textbf{a}_0\equiv(-\frac{k_y}{k^2},\frac{k_x}{k^2})
=\frac{\hat{e}_{\phi}}{k}$ and $[\textbf{k}\times \textbf{A}]_{z}
= -(\lambda^2-\beta^2)/2\gamma^2(\phi)$ which depends only on
$\phi$. After some vector algebra calculations, the spin Hall
conductivity in Eq. (\ref{berrspin}) can be written as
\begin{equation}
\sigma^{s_0}_{xy}=\frac{e}{8\pi^2}\sum_{n}\oint_{C}
d\textbf{l}\cdot
n[\mathbf{k}\times\mathbf{A}]_z\mathbf{a}_0\frac{k^{n}_F}{k^{+}_F-k^{-}_F},
\label{Berrypoten:spin1}
\end{equation}
where $(k^{+}_F-k^{-}_F)=\frac{2m\gamma(\phi)}{\hbar^2}$ was used.
Because $(k^{n}_F-k^{-n}_F)=n(k^{+}_F-k^{-}_F)$, Eq.
(\ref{Berrypoten:spin1}) becomes
\begin{equation}
\sigma^{s_0}_{xy}=\frac{e}{16\pi^2}\sum_{n}\oint_{C}
d\textbf{l}\cdot n[\mathbf{k}\times\mathbf{A}_n]_z\mathbf{a}_0
\label{Berrypoten:spin2}
\end{equation}
where $\mathbf{A}_n=n\mathbf{A}$ with $n=\pm 1$. Using the
definition of Berry phase in Eq. (\ref{Berryphase}), the spin Hall
conductivity can be written as
\begin{equation}
\sigma^{s_0}_{xy}=\frac{e}{16\pi^2}\sum_{n}\oint_{C}
d\textbf{l}\cdot(n\mathbf{A}_n)=\frac{e}{16\pi^2}\sum_{n}n\Phi_n,
\label{Berrypoten:spin3}
\end{equation}
where $\Phi_n=n\Phi$ with $n=\pm 1$, in agreement with Ref.
\onlinecite{she04}. In other words, the spin Hall conductivity can
be written as a line integral of the Berry vector potential or a
surface integral of the Berry curvature. Note that
Eq. (\ref{Berrypoten:spin3}) holds irrespective of the choice of the
eigenstates.

For the OAM Hall conductivity, similarly, Eq. (\ref{berroam})
can be rewritten as
\begin{equation}
\sigma^{o_0}_{xy}=\frac{e}{8\pi^2}\sum_n\oint_{C} d\mathbf{l}\cdot
[\mathbf{k}\times\mathbf{A}_n]^2_{z}\mathbf{a}_0.
\label{Berrypoten:OAM}
\end{equation}
This expression has also been given in Ref. \onlinecite{hu05}, where
%$\mathbf{a}_{0}=\frac{\hat{e}_{\phi}}{k}$ and
%$\mathbf{A}_n=n\mathbf{A}$. In particular, in ~\cite{hu05},
$[\hbar\mathbf{k}\times\mathbf{A}]_{z} =
-\hbar(\lambda^2-\beta^2)/2\gamma^2(\phi)$ was regarded as the OAM
of the eigenstates in the absence of any applied electric field.
We believe that this interpretation is inappropriate because
the expectation value of the conventional OAM operator depends on
the choice of bases, as will be shown in next Subsec. below.
In next Subsec., we also argue that before an
electric field is applied, the OAM of the eigenstates is zero.
Finally, we also find that the ratio of the spin to OAM Hall
conductivity is
$\frac{\sigma^{s_0}_{xy}}{\sigma^{o_0}_{xy}}=2[\textbf{k}\times\textbf{A}]_z|_{\phi=m\pi}$
where $m$ is an integer.

\subsection{Origin of the orbital angular momentum Hall current}

All electrons have an intrinsic spin of $\frac{1}{2}$. Before an
in-plane electric field is applied, the spins of the electrons are
all aligned in the $xy$-plane, as shown in Fig. 1(a)-(b). When an
in-plane electric field is applied, the SO coupling gives rise to
not only a spin transverse force on a moving electron
\cite{li05,nik05} but also an effective SO magnetic torque. The
Rashba-Dresselhaus Hamiltonian can be written as
$(H_R(\mathbf{k})+H_D(\mathbf{k}))=\vec{\sigma}\cdot\mathbf{B}_{eff}$
where
$\mathbf{B}_{eff}\equiv\lambda(\mathbf{k}\times\hat{e}_z)-\beta(k_x\hat{e}_x-k_y\hat{e}_y)$
is the effective SO magnetic field. The dynamics of the z component of
spin can be derived from the Heisenberg equation of motion and we
obtain
$d\sigma_z(t)/dt=\frac{1}{i\hbar}[\sigma_z,\frac{p^2}{2m}+H_D+H_R+eEy]=\frac{1}{i\hbar}[\sigma_z,H_D+H_R]=\frac{1}{i\hbar}[\sigma_z,\sigma_i]B^{i}_{eff}$.
Using the commutation relation of spin matrix
$[\sigma_i,\sigma_j]=2i\epsilon_{ijk}\sigma_k$, we have
$(ds_z(t)/dt)=-\frac{2}{\hbar}(\vec{s}(t)\times\mathbf{B}_{eff})_z$,
where $\vec{s}=\frac{\hbar}{2}\vec{\sigma}$.
%After a short time
%$t_0$, the Fermi surface (i.e., the circle for $n=\pm 1$) would
%move with the distance $y_0=eEt_0/\hbar$.
>From the equation of motion of electron in $k$-space, we also have $\delta
k_y\equiv(k_y(t_0)-k_y(t=0))=-\frac{eE}{\hbar}t_0$ and $\delta
k_x=0$. Therefore, after a short time
$t_0$, the Fermi surface (i.e., the circle for $n=\pm 1$) would
move along -$k_y$ direction with the distance $eEt_0/\hbar$.
This implies that the variation of the effective SO magnetic
field is
$\delta\mathbf{B}_{eff}=(\lambda\hat{e}_x+\beta\hat{e}_y)\delta{k}_y$.
Consequently, each spin feels the effective SO magnetic torque
$\vec{s}(0)\times\delta\mathbf{B}_{eff}$ and tilts out of the $xy$
plane. The quantum dynamical analysis of spin for the Rashba-Dresselhaus
system has been given in~\cite{she04}. Here we
use the quantum perturbation method to evaluate the response
quantities.

Let us expand the wave function to first order of electric field,
$|n\mathbf{k}\rangle'=|n\mathbf{k}\rangle+|n\mathbf{k}\rangle^{(1)}$
where the perturbed wave function is
\begin{equation}
|n\mathbf{k}\rangle^{(1)}=eE\sum_{n'(n'\neq n)}\frac{\langle
n'\mathbf{k}|i\partial/\partial
k_y|n\mathbf{k}\rangle}{E_{n\mathbf{k}}-E_{n'\mathbf{k}}}|n'\mathbf{k}\rangle.
\end{equation}
The expectation value of the $z$ component of spin can be
evaluated by $'\langle n\mathbf{k}|s_z|n\mathbf{k}\rangle'=\langle
s_z\rangle_0+\delta\langle s_z\rangle$ where $\langle
s_z\rangle_0\equiv\langle n\mathbf{k}|s_z|n\mathbf{k}\rangle=0$
and the $z$ component of the spin in first order of electric field
is given by $\delta \langle s_z \rangle \equiv 2Re\langle
n\mathbf{k}|s_z|n\mathbf{k}\rangle^{(1)}=-eE\hbar (\lambda
^2-\beta ^2) \cos \phi /4nk^2 \gamma^3(\phi)$. Therefore, for $k_x>0$, the
spins on the outer (inner) sheet tend to tilt down (up) and for
$k_x<0$, the spins on the outer (inner) sheet tend to tilt up
(down), as shown in the right panel of Fig. 1(b). This results in
transverse spin Hall currents with spin polarization in the $z$
direction.

Now consider the OAM $L_z=\hbar(\mathbf{x}\times\mathbf{k})_z$.
The expectation value $'\langle
n\mathbf{k}|L_z|n\mathbf{k}\rangle'=\langle
L_z\rangle_0+\delta\langle L_z\rangle$ where $\langle
L_z\rangle_0\equiv\langle n\mathbf{k}|L_z|n\mathbf{k}\rangle$ and
$\delta\langle L_z\rangle=2Re\langle
n\mathbf{k}|L_z|n\mathbf{k}\rangle^{(1)}$. In contrast to the spin
case, before an electric field is applied, as argued below, all
the eigenstates carry zero orbital angular momentum, i.e.,
$\langle L_z \rangle _0=0$, as illustrated in the left panel of
Fig. 1(c). In the absence of the applied electric field, the
diagonal matrix elements of the OAM depend on the choice of the
eigenstates, though the off-diagonal matrix elements of the OAM do
not. For example, if we choose the eigenstates of Eq. (5),
together with the conventional position operator $\mathbf{x} =
i\nabla_{\mathbf{k}}$, $\langle L_z \rangle _0=\langle
n\mathbf{k}| \mathbf{x}\times \hbar \mathbf{k} |n\mathbf{k}
\rangle = -\hbar(\lambda ^2-\beta ^2)/2\gamma^2(\phi) $, a value
which has also been obtained in Ref.~\onlinecite{hu05}. On the
other hand, if we use the eigenstates
$|n\mathbf{k}\rangle=\frac{1}{\sqrt{2}}\begin{pmatrix}
e^{-i\theta(\mathbf{k})/2} \\ ine^{i\theta(\mathbf{k})/2}
\end{pmatrix}$, $\langle L_z \rangle _0=0$. This is of course
unsatisfactory, and therefore, we propose to define a gauge
invariant position operator $\mathbf{X} = i\nabla_{\mathbf{k}} +
\mathbf{A}(\mathbf{k})$ to resolve the problem (see Appendix D).
We can show that with this gauge invariant position operator
$\mathbf{X}$, $\langle L_z \rangle _0=0$, irrespective of the
choice of the phase factor of the eigenstates. Therefore, we
believe that all the eigenstates of the Rashba-Dresselhaus system
carry zero OAM in the absence of applied electric fields. Importantly,
the other quantities such as $\delta\langle s_z\rangle$
and $\delta\langle L_z\rangle$, which contain only the $n\neq n'$
matrix elements, are independent of the choice of the phase factor
and also the choice of the position operator.

As an in-plane electric field is applied, the dynamics of the $z$-component
of the OAM can be obtained by means of Heisenberg equation of
motion:
$dL_z(t)/dt=\frac{d}{dt}(\hbar\mathbf{x}\times\mathbf{k})_z=-\frac{2}{\hbar}(\vec{s}(t)\times\mathbf{B}'_{eff})_z+(\mathbf{x}\times\mathbf{F})_z$
where
$\mathbf{B}'_{eff}=\mathbf{B}_{eff}|_{\lambda\rightarrow-\lambda}$
and $\mathbf{F}=-e\mathbf{E}=-eE\hat{e}_y$. The second term
$(\mathbf{x}\times\mathbf{F})_z=\hbar\mathbf{x}\times\dot{\mathbf{k}}$
is the classical external torque which causes the orbital motion
of the electrons and depends on the choice of the origin of the coordinate
system. The first term
$-\frac{2}{\hbar}(\vec{s}\times\mathbf{B}'_{eff})_z$ is the effective
SO magnetic torque. If we use the gauge invariant position
operator, the classical torque $(\mathbf{x}\times\mathbf{F})_z$
does not contribute to the
variation of the OAM in a short time $t_0$. We can show that
$\frac{d}{dt}(\hbar\mathbf{X}\times\mathbf{k})_z=-\frac{2}{\hbar}(\vec{s}(t)\times\mathbf{B}'_{eff})_z-X(t)eE+k_y\Omega_z$,
where
$\Omega_z=\nabla_{\mathbf{k}}\times\mathbf{A}\sim\delta^{(2)}(\mathbf{k})$
and is zero because $\mathbf{k}\neq 0$. After a short
time $t_0$, $\langle X(0)\rangle\delta k_y$ vanishes because
$\langle X(0)\rangle=0$ and the $z$ component of the OAM for each
eigenstate is induced by the effective SO magnetic torque and
external torque only. In our calculations, we treat the potential $eEy$
as a perturbation and use again the quantum
perturbation method instead of solving the Heisenberg equation of
motion. We find that the $z$ component of the OAM is $\delta \langle L_z
\rangle = eE\hbar (\lambda ^2-\beta ^2)^2 \cos \phi /4nk^2
\gamma^5(\phi) = - \delta \langle s_z \rangle [(\lambda ^2-\beta
^2)/\gamma^2(\phi)]$, as illustrated in the right panel of Fig.
1(c).
%Therefore, under the in-plane electric field along the $y$ axis, the electrons
%with $k_x>0$ ($k_x<0$) drift toward the $x$ axis (the $-x$ axis), carrying the downward (upward)
%tilted spinors as well as the upward (downward) OAM with them, and thereby giving rise to
%the OAM Hall current.
Therefore, under the in-plane electric field along the $y$ axis,
the electrons on the outer (inner) sheet with $k_x>0$ drift toward the $x$ axis,
carrying the downward (upward) tilted spinors as
well as finite positive (negative) OAM  $\delta \langle L_z \rangle$, and
the electrons on the outer (inner) sheet with $k_x<0$ drift toward the $x$ axis,
carrying the upward (downward) tilted spinors as
well as finite negative (positive) OAM  $\delta \langle L_z \rangle$. This
gives rise to the OAM Hall current. It is interesting to note
that in the pure Rashba system ($\beta=0$ and
$(\lambda ^2-\beta ^2)/\gamma^2(\phi) = 1$), the $\delta \langle s_z \rangle$ and $\delta \langle
L_z \rangle$ have the same magnitude but the opposite signs
and hence cancel each other exactly. This is  due to the fact
that the $z$-component of the total angular momentum is conserved in
pure Rashba system.

\section{Systems with both Rashba and Dresselhaus spin-orbit couplings}

The calculated total and decomposed spin and OAM Hall conductivities are summarized in Table I.
The total spin and OAM Hall conductivities are displayed as a function of the
ratio $|\lambda/\beta|$ in Fig. 2.
The total angular momentum Hall conductivity
($ \sigma^{s_0}_{xy}+\sigma^{\tau_s}_{xy}+\sigma^{o_0}_{xy}+\sigma^{\tau_o}_{xy}$) is (see Table I)
\begin{equation}
\sigma_{xy}=\left\{ \begin{array}{rcl}
  -\frac{\beta^2}{|\lambda^2-\beta^2|}\frac{e}{4\pi}, & \mbox{} & \lambda^2 \neq \beta^2 \\
         0,               & \mbox{} & \lambda^2=\beta^2      \\
\end{array}\right .
\end{equation}
It is clear that in 2DEG systems with the Rashba-Dresselhaus SO
coupling, the total angular momentum Hall conductivity is in
general not zero, except that $\lambda^2 = \beta^2$ or $\beta=0$
(pure Rashba SO coupling). The sign of the total angular momentum
Hall conductivity is always negative, and, as will be discussed
shortly, this is because the Hall conductivity is dominated by the
negative OAM Hall conductivity. Furthermore, when the two SO
coupling strengths are comparable, the total angular momentum Hall
conductivity is very large, suggesting the interesting possibility
of tuning the angular momentum Hall effect by varying the Rashba
SO coupling strength. This large total Hall conductivity results
from the large OAM Hall conductivity in both the conventional
and present definitions of spin current (see Table I). The conventional
OAM Hall conductivity is the same as that given in Ref. \onlinecite{hu05}.
This singularly large OAM conductivity in the region that $|\lambda/\beta|$
approaches to unity, could lead to spontaneous magnetization,
as suggested in Ref. ~\onlinecite{hu05}. Nevertheless, we believe that this
singular behavior of the OAM conductivity near $\lambda\sim\beta$
is unphysical and is perhaps due to the neglect of the disorder in
the present calculations.
The disorder due to, e.g., impurity scattering and electron-electron
interaction is known to have pronounced effects on the spin Hall
effect ~\cite{ino04,dim05,ocha05}. The infinite large OAM conductivity is expected
to be suppressed by the disorder effects in real 2DEG systems.
Nonetheless, further calculations taking into the disorder effects
are beyond the scope of the present paper.

\begin{table}
\caption{Total and decomposed spin (a) and orbital angular momentum (b) Hall conductivities
of a two-dimensional electron system with Rashba-Dresselhuas spin-orbit coupling.}
\begin{ruledtabular}

\center{(a) Spin Hall conductivity}

\begin{tabular}{|c|c|c|c|}
%\label{tab:spin_conductivity}
%\hline
     & $\lambda^2>\beta^2$ & $\lambda^2=\beta^2  $ &  $\lambda^2<\beta^2$ \\ \hline
$\sigma^{s_0}_{xy}$    & $-\frac{e}{8\pi}$   & $0$   & $\frac{e}{8\pi}$    \\ \hline
$\sigma^{\tau_s}_{xy}$ & $\frac{e}{4\pi}$  & $0$   & $-\frac{e}{4\pi} $  \\ \hline
$\sigma^{s}_{xy}$    & $\frac{e}{8\pi}$   & $0$   & $-\frac{e}{8\pi}$
\end{tabular}

\center{(b) OAM Hall conductivity}

\begin{tabular}{|c|c|c|c|}
%\hline
        & $\lambda^2>\beta^2$ & $\lambda^2=\beta^2  $ &  $\lambda^2<\beta^2$ \\ \hline
$\sigma^{o_0}_{xy}$    & $\frac{\lambda^2+\beta^2}{|\lambda^2-\beta^2|}\frac{e}{8\pi}$   & $0$   & $\frac{\lambda^2+\beta^2}{|\lambda^2-\beta^2|}\frac{e}{8\pi}$    \\ \hline
$\sigma^{\tau_{o}}_{xy}$    & $-\frac{\lambda^2+\beta^2}{|\lambda^2-\beta^2|}\frac{e}{4\pi}$   & $0$   & $-\frac{\lambda^2+\beta^2}{|\lambda^2-\beta^2|}\frac{e}{4\pi}$    \\ \hline
$\sigma^{o}_{xy}$   & $-\frac{\lambda^2+\beta^2}{|\lambda^2-\beta^2|}\frac{e}{8\pi}$   & $0$   & $-\frac{\lambda^2+\beta^2}{|\lambda^2-\beta^2|}\frac{e}{8\pi}$
\end{tabular}

\end{ruledtabular}
\end{table}

Shen~\cite{she04} recently pointed out that in the 2DEG systems
with both Rashba and Dresselhaus couplings, the spin current along
the $z$ direction is antisymmetric with respect to an unitary
transformation: $\sigma_{x}\rightarrow \sigma_{y}$;
$\sigma_{y}\rightarrow \sigma_{x}$; $\sigma_{z}\rightarrow
-\sigma_{z}$. This antisymmetry makes the conventional spin Hall
conductivity changes sign at $\lambda ^2 = \beta ^2$. It is
interesting to note that the spin torque Hall conductivity and
hence the total spin Hall conductivity also change sign at
$\lambda ^2= \beta ^2$ when one moves from the region where
Dresselhaus coupling dominates to the region where Rashba coupling
dominates (Table I and Fig. 2). This shows that our calculated
spin torque and total spin Hall conductivities also obey the
requirement of this antisymmetry. We also note that as for the
case of pure Rashba coupling~\cite{zha05}, the spin torque Hall
conductivity is twice as large as the conventional spin Hall
conductivity but has an opposite sign (Table I), giving rise to
the result that the conserved spin Hall conductivity has the same
size but opposite sign to the conventional spin Hall conductivity.

\begin{figure}
\begin{center}
\includegraphics[width=8cm]{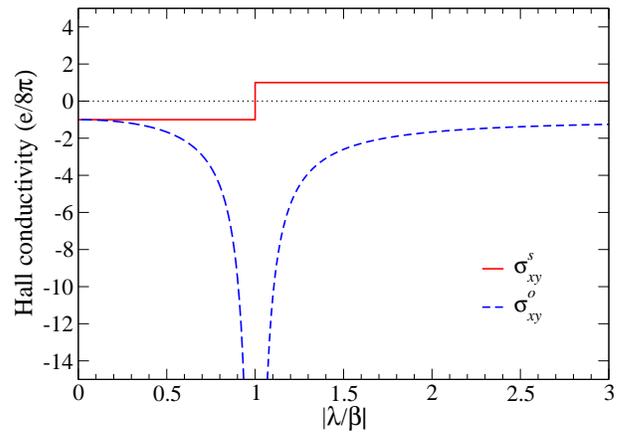}
\end{center}
\caption{(Color online) Spin ($\sigma ^{s} _{xy}$, solid line) and orbital angular momentum
($\sigma^{o}_{xy}$, dashed line) Hall conductivities versus
the strength ratio $|\lambda/\beta|$ of the Rashba ($\lambda$)
to Dresselhaus ($\beta$) spin-orbit coupling.
%The scale of Hall conductivity is $e/8\pi$.
%The Hall conductivities are independent of the carrier density.
%$\sigma ^{s} _{xy}$ is a step function while $\sigma^{OAM}_{xy}$ is diverge near $|\lambda/\beta|=1$.
}
\label{fig2}
\end{figure}

On the other hand, all the (total, torque and conventional) OAM
Hall conductivities do not change sign at $\lambda ^2 = \beta ^2$,
as shown in Table I and Fig. 2. As for the case of spin Hall
effect, the torque OAM Hall conductivity is twice as large as the
conventional OAM Hall conductivity but has an opposite sign,
resulting in that the effective OAM Hall conductivity has the same
size but opposite sign to the conventional OAM Hall conductivity
(Table I). The conventional OAM Hall conductivity has been
reported in Ref.~\onlinecite{hu05}, and our result is consistent
with this previous calculation.

%The relative strength of the Rashba and Dresselhaus SO couplings depends on both the bulk semiconductors
%of which the heterostructure is made and also the strength of the applied gate field.
%The ratio $|\lambda/\beta|=1.5\sim1.7$ in GaAs quantum wells and  $|\lambda/\beta|\sim2.15$
%in InAs quantum well\cite{gan04}.
%By tuning the strength of Rashba coupling $\lambda$, we can control the OAM Hall current and spin Hall current.
%The OAM Hall effect overhelm the spin Hall effect when the two coupling is comparable.

\section{Systems with either pure Rashba term or pure Dresselhaus term}

\subsection{Pure Rashba spin-orbit coupling}
In the case of $\beta=0$ and $\lambda \neq 0$, the Hamiltonian reduces to the pure Rashba Hamiltonian.
The spin torque operator becomes
\begin{equation}
    \tau ^{R}_s\equiv \frac{1}{i\hbar}[s_z,\frac{p^2}{2m}+H_R],
\end{equation}
and the $y$-component of velocity operator is
\begin{equation}
    v^{R}_y=\frac{p_y}{m}+\frac{\lambda}{\hbar}\sigma _x.
\end{equation}
The torque correction term for the spin Hall conductivity is $e/4\pi$, and for the OAM Hall conductivity
is $-e/4\pi$, which exactly cancel each other. This result is confirmed by
the realtion $[L_z,H_R]=-[s_z,H_R]$ that is present in the Rashba Hamiltonian.
Including the torque correction term, we get an effective spin Hall conductivity of $e/8\pi$, and
an effective OAM Hall conductivity of $-e/8\pi$.
This implies that when the torque correction term is taken into account, the spin Hall current is still
exactly canceled by the accompanied OAM Hall current and
there is no total angular momentum current in the pure Rashba system.
This interesting observation has been reported before in Ref. \onlinecite{zha04} in the context
of the conventional definition of the spin and OAM currents.
This may be expected because $[s_z + L_z, H_R] = 0$ and the conservation of the total angular
momentum therefore must be obeyed no matter what definitions of the effective angular momentum
currents one adopts.

Nonetheless, we want to point out here that the spin Hall effect
in the pure Rashba Hamiltonian can still manifest itself and be
detected in several ways, even though the total angular momentum
Hall current is zero. It is well known that the magnetic dipole
moment associated with the spin of an electron is $\mu _s
=-2\mathbf{s}\mu_B$ and the one associated with the OAM of an
electron is $\mu _o =-\mathbf{L}\mu_B$ (see, e.g.,~\cite{sak67}).
Consequently, although the total angular momentum ($s_z+L_z$)
current is zero, the total magnetization [$(\mu^z_s+\mu^z_o) =
-(2s_z + L_z)\mu_B$] current would not be zero and will give rise
to finite magnetization at the edge of the sample. Therefore, the
spin Hall effect in the pure Rashba Hamiltonian can in principle
be probed in at least two ways. As noted in Ref.
\onlinecite{zha04}, one is by measuring the electric field induced
by the nonzero magnetization current~\cite{mei04} and the other is
the magnetization at the sample edges. Recently,
it is suggested in Ref.~\onlinecite{ocha05} that spin current can be
detected by measuring time dependent magnetization precession.

%Because of the spin and OAM Hall current flow in the bulk, the angular momentum might accumulate at the edge of the sample.
%The total angular momentum we consider is the sum of spin and OAM.
%And the magnetic dipole moment induced by electron spin is\cite{sak67}:
%\begin{equation}
%   \mu _e =-\frac{ge}{m}\mathbf{S},
%\end{equation}
%which has a $g$ factor of 2.
%However, the magnetic dipole moment contributed from the OAM is
%\begin{equation}
%   \mu _L=-\frac{e}{m}\mathbf{L}.
%\end{equation}
%The magnetization accumulation at the edge of the smaple contributed from spin and OAM Hall current might be finite in a pure Rashba system.

\subsection{Pure Dresselhaus spin-orbit coupling}
In the case of $\lambda=0$ and $\beta \neq 0$, the Hamiltonian reduces to the pure Dresselhaus
Hamiltonian.
In this case, the total, torque correction and conventional spin and OAM Hall conductivities have
the same signs (Table I). For example, the total spin Hall conductivity is  $-e/8\pi$, being the same
as the total OAM Hall conductivity, thereby giving rise to a total angular momentum Hall
conductivity of $-e/4\pi$.
In the Dresselhaus Hamiltonian, the total angular momentum is not conserved, i.e.,
$[s_z + L_z, H_D] = -2i\beta(\sigma_x p_y + \sigma_y p_x) \neq 0$.
Instead, we find $[s_z, H_D]=[L_z, H_D]$ or $[s_z - L_z, H_D] = 0$.
Therefore, the spin and OAM Hall conductivities would add up rather than
cancel each other, in contrast to the Rashba Hamiltonian.

We notice that the spin Hall conductivity in the clean limit is
constant in either pure Rashba system, or pure Dresselhaus system,
or the mixture Rashba-Dresselhaus system $[$see Table I(a)$]$.
Interestingly, in Refs. ~\onlinecite{dim05} and ~\onlinecite{ocha05}, 
it is pointed that
the constant spin Hall conductivity in pure Rashba system would
result in an unphysically growth of the magnetization with time.
This is due to the fact that the conventionally defined spin Hall
current in the pure Rashba system turns out to be proportional to the
time derivative of the magnetizaion of the system. This linear
relation between the conventional spin Hall current and the time
derivative of the magnetization should also exist in the pure
Dresselhaus system because the two systems are related by the
unitary transformation: $\sigma_x\rightarrow\sigma_y$ and
$\sigma_y\rightarrow\sigma_x$, as mentioned above in Sec. III.
Indeed, we find that the conventional spin current in the pure
Dresselhaus system is
$j^{s_z}_x=\frac{\hbar}{4}\{v_x,\sigma_z\}=\frac{\hbar}{2m}\sigma_zp_x$,
where the relation $\{\sigma_i,\sigma_j\}=2\delta_{ij}$ was used.
Furthermore, it can be shown that the spin precession and
conventional spin current satisfy the relations:
$\frac{d\langle\sigma_y(t)\rangle}{dt}=\frac{2\beta}{\hbar^2}\langle
p_x\sigma_z\rangle(t)=\frac{4m\beta}{\hbar^3}\langle
j^{s_z}_x(t)\rangle$ and
$\frac{d\langle\sigma_x(t)\rangle}{dt}=\frac{4m\beta}{\hbar^3}\langle
j^{s_z}_y(t)\rangle$. Therefore, the constant
conventional spin current in the clean limit in the pure
Dresselhaus
%absence of disorder is found
%to be the constant $\frac{e}{8\pi}E_y$ (and $-\frac{e}{8\pi}E_x$)
system would also lead to the unphysical consequence
that both $\langle\sigma_y(t)\rangle$ and
$\langle\sigma_x(t)\rangle$ depend linearly on time, resulting
in an infinite growth of the magnetization.
In contrast, because the conserved spin current satisfies the
continuity equation $[$Eq. (1)$]$, we find no such relation that
the conserved spin current is proportional to the time derivative
of the spin operator, thereby, free from the artifact discussed above,
as also pointed out in Ref. ~\onlinecite{Sugi06}.

%In this paper, we do not consider the disorder
%effect of Rashba-Dresselhaus system with the definition of
%conserved spin current. Nevertheless, in the $k$-linear and
%$k$-cubic Rashba system, the disorder effect with the definition
%of conserved spin current has been studied in~\cite{Sugi06}.

%Thus, the Dresselhaus term plays an important role on the non-zero magnetization current.
%The physical origin is the total angular momentum is not conserved when the system is involved with the Dresselhaus SO coupling.
%We can see that the commutation relation $[J_z,H_0]=[s_z+L_z,H_D]=-2i\beta(\sigma _x p_y +\sigma _y p_x)$ is not zero.

\section{Conclusions}
In conclusion, we have calculated the spin Hall coductivity in the
Rashba-Dresselhaus Hamiltonian with the spin torque correction in
the absence of disorder, and find it to have the same magnitude
but an opposite sign to the result reported
before~\cite{sin04,she04}. The spin Hall conductivity in the
absence of disorder is still a constant in pure Dresselhaus ( or
Rashba ) system even when the torque correction is considered.
We also introduce the conserved
effective OAM current and find that in general, the OAM
Hall effect does not cancel the spin Hall effect in the 2DEG with
the Rashba-Dresselhaus spin-orbit coupling. The OAM Hall
conductivity depends significantly on the strength ratio of the
Rashba to Dresselhaus spin-orbit coupling, suggesting that one can
manipulate the total Hall current through tuning the Rashba
coupling by a gate voltage~\cite{Schl03,gani04}.
%However, the
%unexpected large value of OAM Hall conductivity appears in this 2D
%system as the two strength are comparable. The huge OAM
%conductivity in the region that $|\lambda/\beta|$ approaches to
%unity would lead to spontaneous magnetization~\cite{hu05}.
We argue that in a pure Rashba system, though the spin Hall
conductivity is exactly cancelled by the OAM Hall conductivity due
to the angular momentum conservation, the spin Hall effect still
manifest itself as nonzero magnetization Hall current and finite
magnetization at the sample edges because the magnetic dipole
moment associated with the spin of an electron is twice as large
as that of the OAM. We show that the spin and OAM Hall
conductivities have a simple relation to the Berry vector (or
gauge) potential. We also calculate the electric field-induced OAM
and discuss the origin of the OAM Hall current.
%In the case of $|\lambda|=|\beta|$, the OAM Hall conductivity we get is zero.

\section*{Acknowledgments}
The authors thank Ming-Che Chang and Qian Niu for stimulating discussions.
The authors gratefully acknowledge financial supports from the National Science Council
of Taiwan.

%\textbf{Appendix} : \textbf{Details of calculation of torque}\\
%\\
%\begin{table}
%\begin{tabular}{|c||c|c|c|}
%\hline
%        & $\lambda^2>\beta^2$ & $\lambda^2=\beta^2  $ &  $\lambda^2<\beta^2$ \\
%\hline
%$\sigma^{s}_{xy}$    & $-\frac{e}{8\pi}$   & $0$   & $\frac{e}{8\pi}$    \\
%\hline\lv
%$\sigma^{\tau_s}_{xy}$ & $\frac{e}{4\pi}$  & $0$   & $-\frac{e}{4\pi} $  \\
%\hline
%$\sigma^{s_z}_{xy}$    & $\frac{e}{8\pi}$   & $0$   & $-\frac{e}{8\pi}$    \\
%\hline
%\end{tabular}
%\caption{ The list of spin current conductivity in 2DES system
%with Rashba and Dresselhuas coupling.}
%\label{tab:spin_conductivity}
%\end{table}
%\begin{table}
%\begin{tabular}{|c||c|c|c|}
%\hline
%        & $\lambda^2>\beta^2$ & $\lambda^2=\beta^2  $ &  $\lambda^2<\beta^2$ \\
%\hline
%$\sigma^{OAM}_{xy}$    & $\frac{\lambda^2+\beta^2}{\lambda^2-\beta^2}\frac{e}{8\pi}$   & $0$   & $-\frac{\lambda^2+\beta^2}{\lambda^2-\beta^2}\frac{e}{8\pi}$    \\
%\hline
%$\sigma^{\tau_{OAM}}_{xy}$    & $-\frac{\lambda^2+\beta^2}{\lambda^2-\beta^2}\frac{e}{4\pi}$   & $0$   & $\frac{\lambda^2+\beta^2}{\lambda^2-\beta^2}\frac{e}{4\pi}$    \\
%\hline
%$\sigma^{L_z}_{xy}$    & $-\frac{\lambda^2+\beta^2}{\lambda^2-\beta^2}\frac{e}{8\pi}$   & $0$   & $\frac{\lambda^2+\beta^2}{\lambda^2-\beta^2}\frac{e}{8\pi}$    \\
%\hline
%\end{tabular}
%\caption{ The list of OAM current conductivity in 2DES system with
%Rashba and Dresselhuas coupling.} \label{tab:OAM_conductance}
%\end{table}
\appendix

\section{The continuity equation for the effective spin current}
In this appendix, we derive the continuity equation for the
effective spin current. A similar derivation can also be found
in~\cite{Sun05}. The definition of spin density is
$\mathcal{S}_z=\Psi^{\dag}s_z\Psi$, where
$s_z=\frac{\hbar}{2}\sigma_z$. Using
$i\hbar\frac{\partial}{\partial t}\Psi=H\Psi$ and
$H=H_0+e\mathbf{E}\cdot\mathbf{x}$, we obtain the following
equation:
\begin{equation}
\begin{split}
\frac{\partial}{\partial t}[\Psi^{\dag}\sigma_{z}\Psi]
=&\frac{1}{i\hbar}(\Psi^{\dag}\sigma_zH_0\Psi-(H_0\Psi)^{\dag}\sigma_z\Psi)\\
=&\frac{1}{i\hbar}\{[\Psi^{\dag}\sigma_{z}\frac{p^2}{2m}\Psi-(\frac{p^2}{2m}\Psi)^{\dag}\sigma_z\Psi]\\
&+[\Psi^{\dag}\sigma_{z}H_R\Psi-(H_R\Psi)^{\dag}\sigma_z\Psi]\\
&+[\Psi^{\dag}\sigma_{z}H_D\Psi-(H_D\Psi)^{\dag}\sigma_z\Psi]\},
\label{conspin1}
\end{split}
\end{equation}
where we have used $[s_z,\mathbf{x}]=0$. First, we combine the
following two equations:
\begin{equation*}
\frac{\partial}{\partial x_i}(\Psi^{\dag}\frac{\partial}{\partial
x_i}\sigma_z\Psi)=(\frac{\partial\Psi^{\dag}}{\partial
x_i})\frac{\partial}{\partial
x_i}\sigma_z\Psi+\Psi^{\dag}\frac{\partial}{\partial
x_i}\frac{\partial}{\partial x_i}\sigma_z\Psi
\end{equation*}
\begin{equation*}
\frac{\partial}{\partial x_i}(\frac{\partial\Psi^{\dag}}{\partial
x_i}\sigma_z\Psi)=(\frac{\partial}{\partial
x_i}\frac{\partial}{\partial
x_i}\Psi^{\dag})\sigma_z\Psi+\frac{\partial\Psi^{\dag}}{\partial
x_i}(\frac{\partial}{\partial x_i}\sigma_z\Psi).
\end{equation*}
Using the definition of momentum operator
$\mathbf{p}=-i\hbar\nabla$ and
$(\Psi^{\dag}\mathbf{p}\sigma_z\Psi)^{\dag}=-(\mathbf{p}\Psi^{\dag})\sigma_z\Psi$,
we obtain :
\begin{equation}
\mathbf{p}\cdot[2Re(\Psi^{\dag}\mathbf{p}\sigma_z\Psi)]=[\Psi^{\dag}p^2\sigma_z\Psi-(p^2\Psi^{\dag})\sigma_z\Psi].
\label{conspin:mon}
\end{equation}
Next, consider the term
$[\Psi^{\dag}\sigma_zH_R\Psi-(H_R\Psi)^{\dag}\sigma_z\Psi]$ which
contains only the Rashba Hamiltonian. We obtain the following
equations:
\begin{equation}
\begin{split}
&[\Psi^{\dag}\sigma_zH_R\Psi-(H_R\Psi)^{\dag}\sigma_z\Psi]\\
&=\frac{2i\lambda}{\hbar}\Psi^{\dag}\mathbf{p}\cdot\vec{\sigma}\Psi-\frac{\lambda}{\hbar}\mathbf{p}\cdot[\Psi^{\dag}\sigma_z(\vec{\sigma}\times\hat{e}_z)\Psi]
\label{conspin:R1}
\end{split}
\end{equation}
and
\begin{equation}
\begin{split}
&[\Psi^{\dag}\sigma_zH_R\Psi-(H_R\Psi)^{\dag}\sigma_z\Psi]\\
&=-[\Psi^{\dag}\sigma_zH_R\Psi-(H_R\Psi)^{\dag}\sigma_z\Psi]^{\dag}\\
&=\frac{2i\lambda}{\hbar}[\Psi^{\dag}\mathbf{p}\cdot\vec{\sigma}\Psi]^{\dag}-\frac{\lambda}{\hbar}\mathbf{p}\cdot[\Psi^{\dag}\sigma_z(\vec{\sigma}\times\hat{e}_z)\Psi]^{\dag}.
\label{conspin:R2}
\end{split}
\end{equation}
We combine Eq. (\ref{conspin:R1}) and Eq. (\ref{conspin:R2}):
\begin{equation}
\begin{split}
&[\Psi^{\dag}\sigma_zH_R\Psi-(H_R\Psi)^{\dag}\sigma_z\Psi]\\
&=\frac{2i\lambda}{\hbar}Re[\Psi^{\dag}(\mathbf{p}\cdot\vec{\sigma})\Psi]-\frac{\lambda}{\hbar}\mathbf{p}\cdot
Re[\Psi^{\dag}\sigma_z(\vec{\sigma}\times\hat{e}_z)\Psi].
\label{conspin:R3}
\end{split}
\end{equation}
Now consider the term :
$[\Psi^{\dag}\sigma_{z}H_D\Psi-(H_D\Psi)^{\dag}\sigma_z\Psi]$
which contains only the Dresselhaus Hamiltonian. From the second term
$(H_D\Psi)^{\dag}\sigma_z\Psi$, we obtain:
\begin{equation}
\begin{split}
(H_D\Psi)^{\dag}\sigma_z\Psi
=&-\frac{\beta}{\hbar}[-p_x\Psi^{\dag}\sigma_x+p_y\Psi^{\dag}\sigma_y]\sigma_z\Psi\\
=&~\frac{\beta}{\hbar}p_x(\Psi^{\dag}\sigma_x\sigma_z\Psi)-\frac{\beta}{\hbar}\Psi^{\dag}p_x\sigma_x\sigma_z\Psi\\
&-\frac{\beta}{\hbar}p_y(\Psi^{\dag}\sigma_y\sigma_z\Psi)+\frac{\beta}{\hbar}\Psi^{\dag}p_y\sigma_y\sigma_z\Psi\\
=&-\frac{\beta}{\hbar}\{p_x(\Psi^{\dag}\sigma_z\sigma_x\Psi)-p_y(\Psi^{\dag}\sigma_z\sigma_y\Psi)\}\\
&-\frac{\beta}{\hbar}\Psi^{\dag}(p_x\sigma_x-p_y\sigma_y)\sigma_z\Psi.
\end{split}
\end{equation}
and then:
\begin{equation}
\begin{split}
&[\Psi^{\dag}\sigma_{z}H_D\Psi-(H_D\Psi)^{\dag}\sigma_z\Psi]\\
=&\frac{-2i\beta}{\hbar}\Psi^{\dag}(p_x\sigma_y+p_y\sigma_x)\Psi\\
&+\frac{\beta}{\hbar}\{p_x[\Psi^{\dag}\sigma_z\sigma_x\Psi]-p_y[\Psi^{\dag}\sigma_z\sigma_y\Psi]\}\\
=&\frac{-2i\beta}{\hbar}Re[\Psi^{\dag}(p_x\sigma_y+p_y\sigma_x)\Psi]\\
&+\frac{\beta}{\hbar}\{p_xRe[\Psi^{\dag}\sigma_z\sigma_x\Psi]-p_yRe[\Psi^{\dag}\sigma_z\sigma_y\Psi]\},
\label{conspin:D3}
\end{split}
\end{equation}
where we have used the relation:
$[\Psi^{\dag}\sigma_zH_D\Psi-(H_D\Psi)^{\dag}\sigma_z\Psi]=-[\Psi^{\dag}\sigma_zH_D\Psi-(H_D\Psi)^{\dag}\sigma_z\Psi]^{\dag}$.
Substituting Eqs. (\ref{conspin:mon}), (\ref{conspin:R3}) and
(\ref{conspin:D3}) into (\ref{conspin1}), we obtain
\begin{equation}
\begin{split}
&\frac{\partial}{\partial t}[\Psi^{\dag}\sigma_{z}\Psi]\\
=&-\nabla\cdot
Re[\Psi^{\dag}\sigma_z(\frac{\mathbf{p}}{m}-\frac{\lambda}{\hbar}\vec{\sigma}\times\hat{e}_z-\frac{\beta}{\hbar}(\sigma_x\hat{e}_x-\sigma_y\hat{e}_y))\Psi]\\
&+\frac{2\lambda}{\hbar^2}Re[\Psi^{\dag}(\mathbf{p}\cdot\vec{\sigma})\Psi]-\frac{2\beta}{\hbar^2}Re[\Psi^{\dag}(p_x\sigma_y+p_y\sigma_x)\Psi].
\label{conspin2}
\end{split}
\end{equation}
Consider the commutator $[\sigma_z,H_0]$, and we have
\begin{equation}
\frac{1}{i\hbar}[\sigma_z,\frac{p^2}{2m}+H_R+H_D]=\frac{2\lambda}{\hbar^2}\mathbf{p}\cdot\vec{\sigma}+\frac{-2\beta}{\hbar^2}(p_x\sigma_y+p_y\sigma_x).
\label{torquespin}
\end{equation}
The velocity operator in the Rashba-Dresselhaus system satisfy the
following relation:
\begin{equation}
\begin{split}
\mathbf{v}&=\frac{1}{i\hbar}[\textbf{x},H]=\frac{1}{i\hbar}[\textbf{x},H_0]\\
&=\frac{\textbf{p}}{m}-\frac{\lambda}{\hbar}(\vec{\sigma}\times\hat{e}_z)-\frac{\beta}{\hbar}(\sigma_x\hat{e}_x-\sigma_y\hat{e}_y).
\label{velocity}
\end{split}
\end{equation}
Finally, substituting Eq. (\ref{velocity}) and Eq.
(\ref{torquespin}) into Eq. (\ref{conspin2}), we obtain the continuity
equation of the effective spin current
\begin{equation}
\frac{\partial}{\partial t}[\Psi^{\dag}s_z\Psi]=-\nabla\cdot
Re[\Psi^{\dag}\mathbf{j}^{s_z}\Psi]+Re[\Psi^{\dag}\tau_s\Psi],
\label{conspin3}
\end{equation}
where we have used $s_z=\frac{\hbar}{2}\sigma_z$,
$\mathbf{j}^{s_z}=\frac{1}{2}\{\mathbf{v},s_z\}$ and
$\tau_{s}=\frac{1}{i\hbar}[s_z,H_0]$. With the definition of spin
density $\mathcal{S}_z=\Psi^{\dag}s_z\Psi$, spin current
$\mathbf{J}_s=Re[\Psi^{\dag}\mathbf{j}^{s_z}\Psi]$ and spin torque
$\mathcal{T}^{s}_{z}=Re[\Psi^{\dag}\tau_s\Psi]$, Eq.
(\ref{conspin3}) can be rewritten as
\begin{equation}
\frac{\partial\mathcal{S}_z}{\partial
t}+\nabla\cdot\mathbf{J}_s=\mathcal{T}^{s}_{z}
\end{equation}
which is in agreement with Ref. \onlinecite{zha05}.

\section{The continuity equation for the effective OAM current}
In this appendix, we derive the continuity equation for the
effective OAM current. Let us consider
the $z$-component of the OAM density $\mathcal{L}_z\equiv\Psi^{\dag}L_{z}\Psi$
and its partial time-derivative $i\hbar\frac{\partial}{\partial
t}\mathcal{L}_z$. Applying $i\hbar\frac{\partial}{\partial
t}\Psi=H\Psi$ where $H=H_0+e\mathbf{E}\cdot\mathbf{x}$, we have
\begin{equation}
\begin{split}
&i\hbar\frac{\partial}{\partial t}(\Psi^{\dag}L_{z}\Psi)\\
=&~\Psi^{\dag}L_{z}H_0\Psi-(H_0\Psi)^{\dag}L_{z}\Psi+\Psi^{\dag}[L_z,e\mathbf{E}\cdot\mathbf{x}]\Psi\\
=&~\Psi^{\dag}L_{z}(\frac{p^2}{2m}+H_{R}+H_{D})\Psi-[(\frac{p^2}{2m}+H_{R}+H_{D})\Psi]^{\dag}L_{z}\Psi\\
&+\Psi^{\dag}[L_z,e\mathbf{E}\cdot\mathbf{x}]\Psi
\label{OAMcon}
\end{split}
\end{equation}
First, using the commutation relations
$[\textbf{p},L_z]=i\hbar\hat{e}_z\times\textbf{p}$ and
$[L_z,p^2]=0$, we can show that:
\begin{equation}
\begin{split}
&\{\Psi^{\dag}L_{z}p^2\Psi-(p^2\Psi^{\dag})L_{z}\Psi\}\\
=&~\textbf{p}\cdot[\Psi^{\dag}\textbf{p}L_z\Psi-(\textbf{p}\Psi^{\dag})(L_z\Psi)]\\
=&~\textbf{p}\cdot[\Psi^{\dag}\{\textbf{p},L_{z}\}\Psi]
-p^2(\Psi^{\dag}L_{z}\Psi)+\textbf{p}\cdot[\Psi^{\dag}i\hbar\hat{e}_{z}\times
\textbf{p}\Psi].\\
\label{OAMcon1}
\end{split}
\end{equation}
%by using the commutation relations
%$[\textbf{p},L_z]=i\hbar\hat{e}_z\times\textbf{p}$ and
%$[L_z,p^2]=0$.
For the Rashba Hamiltonian $H_R$, we have
\begin{eqnarray*}
(H_R\Psi)^{\dag}L_z\Psi&=&\frac{\lambda}{\hbar}[-p_y\Psi^{\dag}\sigma_x+p_x\Psi^{\dag}\sigma_y]L_z\Psi\\
                      &=&\frac{\lambda}{\hbar}\textbf{p}\cdot[\Psi^{\dag}(\vec{\sigma}\times\hat{e}_z)L_z\Psi]\\
                     &&+\frac{\lambda}{\hbar}\Psi^{\dag}(p_y\sigma_x-p_x\sigma_y)L_z\Psi
\end{eqnarray*}
and
\begin{equation}
\begin{split}
&\{\Psi^{\dag}L_{z}H_{R}\Psi-(H_{R}\Psi)^{\dag}L_{z}\Psi\}\\
&=\Psi^{\dag}[L_{z},H_{R}]\Psi-\frac{\lambda}{\hbar}\textbf{p}\cdot[\Psi^{\dag}(\vec{\sigma}\times\hat{e}_{z})L_z\Psi].
\label{OAMcon2}
\end{split}
\end{equation}
For the Dresselhaus Hamiltonian $H_D$, we obtain
\begin{equation}
\begin{split}
&\{\Psi^{\dag}L_{z}H_{D}\Psi-(H_{D}\Psi)^{\dag}L_{z}\Psi\}\\
&=\Psi^{\dag}[L_{z},H_{D}]\Psi-\frac{\beta}{\hbar}\textbf{p}\cdot[\Psi^{\dag}(\sigma_{x}\hat{e}_{x}-\sigma_{y}\hat{e}_{y})L_z\Psi].
\label{OAMcon3}
\end{split}
\end{equation}
Substituting Eqs. (\ref{OAMcon1}), (\ref{OAMcon2}) and
(\ref{OAMcon3}) into Eq. (\ref{OAMcon}), we have
\begin{equation}
\begin{split}
&i\hbar\frac{\partial}{\partial t}(\Psi^{\dag}L_{z}\Psi)\\
=&~\textbf{p}\cdot[\Psi^{\dag}\{\frac{\textbf{p}}{2m},L_{z}\}\Psi+\Psi^{\dag}[L_{z},e\mathbf{E}\cdot\mathbf{x}]\Psi\\
&-\frac{\lambda}{\hbar}\Psi^{\dag}(\vec{\sigma}\times\hat{e}_{z})L_z\Psi-\frac{\beta}{\hbar}\Psi^{\dag}(\sigma_{x}\hat{e}_{x}-\sigma_{y}\hat{e}_{y})L_z\Psi]\\
&+\Psi^{\dag}[L_{z},H_{R}+H_{D}]\Psi-\frac{1}{2m}p^2(\Psi^{\dag}L_{z}\Psi)\\
&+\frac{1}{2m}\textbf{p}\cdot[\Psi^{\dag}i\hbar\hat{e}_{z}\times
\textbf{p}\Psi]. \label{OAMcon4}
\end{split}
\end{equation}

Putting Eq. (\ref{velocity}) into Eq. (\ref{OAMcon4}), we have
\begin{eqnarray*}
\begin{split}
&\frac{\partial}{\partial
t}(\Psi^{\dag}L_{z}\Psi)\\
=&-\nabla\cdot[\Psi^{\dag}\frac{1}{2}\{\textbf{v},L_z\}\Psi]+\Psi^{\dag}\frac{1}{i\hbar}[L_{z},e\mathbf{E}\cdot\mathbf{x}]\Psi\\
&+\Psi^{\dag}\frac{1}{i\hbar}[L_{z},H_0]\Psi-\frac{i\hbar}{2m}\nabla^{2}(\Psi^{\dag}L_{z}\Psi)-\textbf{p}\cdot[\Psi^{\dag}\hat{e}_{z}\times
\textbf{p}\Psi]
\end{split}
\end{eqnarray*}
Finally, it can be shown that
$\textbf{p}\cdot[\Psi^{\dag}\hat{e}_{z}\times \textbf{p}\Psi]$ is
purely imaginary, and $\Psi^{\dag}L_z\Psi =Re(\Psi^{\dag}L_z\Psi)$ is real
for both the eigenstates of the Rashba-Dresselhauls Hamiltonian and the
Bloch states $\Psi_{\textbf{k}}$.
%namely, $\Psi^{\dag}L_z\Psi=Re(\Psi^{\dag}L_z\Psi)$.
The $\mathcal{L}_z$ continuity equation becomes :
\begin{equation}
\frac{\partial}{\partial t}(\Psi^{\dag}L_{z}\Psi)=-\nabla\cdot
Re[\Psi^{\dag}\mathbf{j}^{o_0}\Psi]+Re[\Psi^{\dag}\tau_{o}\Psi]+Re[\Psi^{\dag}\tau_{E}\Psi]
\label{OAMcon5}
\end{equation}
where $\mathbf{j}^{o_0}=\frac{1}{2}\{\textbf{v},L_{z}\}$ ,
$\tau_{o}=\frac{1}{i\hbar}[L_{z},H_0]$ and
$\tau_E=\frac{1}{i\hbar}[L_{z},e\mathbf{E}\cdot\mathbf{x}]$. With
the OAM current density
$\mathbf{J}_{o}=Re[\Psi^{\dag}\mathbf{j}^{o_0}\Psi]$, the OAM torque density
$\mathcal{T}^{o}_{z}=Re[\Psi^{\dag}\tau_{o}\Psi]$ and the classical
torque density $\mathcal{T}^{E}_z=Re[\Psi^{\dag}\tau_{E}\Psi]$, Eq.
(\ref{OAMcon5}) can be rewritten as
\begin{equation}
\frac{\partial\mathcal{L}_z}{\partial
t}+\nabla\cdot\mathbf{J}_{o}=\mathcal{T}^{o}_{z}+\mathcal{T}^{E}_{z},
\label{OAMcon6}
\end{equation}
which is the OAM continuity equation.

The torque $\mathcal{T}^{E}_z=Re[\Psi^{\dag}\tau_{E}\Psi]$ has a
classical analogue. It can be regarded as the rotational torque moment due to the force
$e\mathbf{E}$ exerted on a particle located at the position $\mathbf{x}$
with respect to the origin of the coordinate system. This can be seen from the
commutator $[L_z,e\mathbf{E}\cdot\mathbf{x}]=-i\hbar
e(\mathbf{x}\times\mathbf{E})_z$, and
\begin{equation}
\mathcal{T}^{E}_z=-Re[(\Psi^{\dag}\mathbf{x}\Psi)\times
e\mathbf{E}]_z.
\end{equation}
>From the space integral $\int
dV\mathcal{T}^{E}_z=-Re[(\int dV\Psi^{\dag}\mathbf{x}\Psi)\times
e\mathbf{E}]_z$, we obtain $\int
dV\mathcal{T}^{E}_z=-Re[\langle\Psi|\mathbf{x}|\Psi\rangle\times(e\mathbf{E})]_z$,
where $\int
dV\Psi\mathbf{x}\Psi=\langle\Psi|\mathbf{x}|\Psi\rangle$.
%The expectation value $\langle\Psi|\mathbf{x}|\Psi\rangle$
%represents the position of wave packet with respect to the origin
%of coordinate system.
Expanding the
$\langle\Psi|\mathbf{x}|\Psi\rangle$ in powers of electric field,
we have
$\langle\Psi|\mathbf{x}|\Psi\rangle=\langle\Psi_0|\mathbf{x}|\Psi_0\rangle+\langle\mathbf{x}\rangle^{(1)}+o(E^2)$,
where $|\Psi_0\rangle$ satisfies the unperturbed wave equation
$i\hbar\frac{\partial}{\partial
t}|\Psi_0\rangle=H_0|\Psi_0\rangle$. Therefore, we obtain $\int
dV\mathcal{T}^{E}_z=-Re[\langle\Psi_0|\mathbf{x}|\Psi_0\rangle\times(e\mathbf{E})]_z+o(E^2)$.
However, we should demand that the expectation value of
the OAM in the unperturbed system is zero. This would imply that the
expectation value of the position operator must be zero in the absence
of the external electric field, as discussed in Appendix D. Finally, we
obtain that the OAM continuity equation
\begin{equation}
\frac{\partial\mathcal{L}_z}{\partial
t}+\nabla\cdot\mathbf{J}_{o}=\mathcal{T}^{o}_{z}
\end{equation}
which is valid in first order of the electric field.

\section{Spin and OAM torque Hall conductivity}
In this appendix, we derive the spin and orbital angular momentum
(OAM) torque Hall conductivities ($\sigma^{\tau}_{xy}$)
in the Rashba-Dresselhaus system. The conductivity in pure Rashba
system can be obtained by setting $\beta=0$ in the
$\lambda^2>\beta^2$ condition. The total torque conductivity
is given by
\begin{equation}
\sigma^{\tau}_{xy}=(\sigma^{\tau_s}_{xy}+\sigma^{\tau_{o}}_{xy})=Re[i\partial\chi_y(\mathbf{q})/\partial
q_x|_{q_{x}\rightarrow 0}],
\label{tor1}
\end{equation}
where $\chi_y(q)$ is defined as
\begin{equation}
   \begin{split}
    \mathbf{\chi}_y(\mathbf{q})&=\frac{ie\hbar}{V}\sum_{n \neq n'}\sum_{\mathbf{k}}[f_{n\mathbf{k}}-f_{n'\mathbf{k}+\mathbf{q}}]\\
                 &\times
                 \frac{\langle n\mathbf{k}|\tau(\mathbf{k},\mathbf{q})|n'\mathbf{k}+\mathbf{q}\rangle\langle
                 n'\mathbf{k}+\mathbf{q}|v_y(\mathbf{k},\mathbf{q})|n\mathbf{k}\rangle}{(E_{n\mathbf{k}}-E_{n'\mathbf{k}+\mathbf{q}})^2},
                 \label{tor2}
    \end{split}
\end{equation}
where
$\tau(\mathbf{k},\mathbf{q})=\frac{1}{2}[\tau(\mathbf{k})+\tau(\mathbf{k}+\mathbf{q})]$,
$\mathbf{v}(\mathbf{k},\mathbf{q})=\frac{1}{2}[\mathbf{v}(\mathbf{k})+\mathbf{v}(\mathbf{k}+\mathbf{q})]$
and $\tau=\frac{1}{i\hbar}[s_z+L_z,H_0]$ is the total torque. The
equation (\ref{tor1}) suggests that we can choose
$\mathbf{q}=(q_x,0,0)$ for simplicity. By the choice
of $\mathbf{q}=(q_x,0,0)$, the $\chi_y(\mathbf{q})$ can be
expanded in power of $q_x$, i.e.,
$\chi_y(\mathbf{q})=\chi_y^{(0)}+\chi^{(1)}_yq_x+\chi_y^{(2)}q_x^2+o(q_x^3)$,
and the torque conductivity will be rewritten as:
$\sigma^{\tau}_{xy}=Re[i\chi^{(1)}_y]=-Im[\chi^{(1)}_y]$ by using
Eq. (\ref{tor1}).

To evaluate $\chi_y$ in Eq. (\ref{tor2}), we first expand the
matrix elements $\langle
n\textbf{k}|\tau(\textbf{k},\textbf{q})|n'\textbf{k}+\textbf{q}\rangle\langle
n'\textbf{k}+\textbf{q}|v(\textbf{k},\textbf{q})_y|n\textbf{k}\rangle$
to first order of $q_x$:
\begin{equation}
\begin{split}
&\langle
n\textbf{k}|\tau(\textbf{k},\textbf{q})|n'\textbf{k}+\textbf{q}\rangle\langle
n'\textbf{k}+\textbf{q}|v(\textbf{k},\textbf{q})_y|n\textbf{k}\rangle\\
=&-\frac{n\beta\hbar}{m}k_y\frac{\partial\theta}{\partial
k_x}(k_x\sin\theta+k_y\cos\theta)q_x\\
&-\frac{\lambda\beta}{\hbar}[k_y+k_y\cos(2\theta)+k_x\sin(2\theta)\\
&+q_x\frac{\partial\theta}{\partial
k_x}(k_x\cos(2\theta)-k_y\sin(2\theta))+\frac{q_x}{2}\sin(2\theta)]\\
&-\frac{\beta^2}{\hbar}[k_x+k_y\sin(2\theta)-k_x\cos(2\theta)\\
&+q_x\frac{\partial\theta}{\partial
k_x}(k_y\cos(2\theta)+k_x\sin(2\theta))\\
&+\frac{q_x}{2}(1-\cos(2\theta))]+o(q_x^2).\\
\label{tor3}
\end{split}
\end{equation}
We also need to expand the
$(f_{n\mathbf{k}}-f_{n'\mathbf{k}+\mathbf{q}})/(E_{n\mathbf{k}}-E_{n'\mathbf{k}+\mathbf{q}})^{2}$
to first order of $q_x$. We have
\begin{equation}
\begin{split}
\frac{f_{n\mathbf{k}}-f_{n'\mathbf{k}+\mathbf{q}}}{(E_{n\mathbf{k}}-E_{n'\mathbf{k}+\mathbf{q}})^{2}}
=&\frac{f_{n\mathbf{k}}-f_{n'\mathbf{k}}}{(E_{n\mathbf{k}}-E_{n'\mathbf{k}})^{2}}\\
&+q_x\frac{\partial E_{n'\mathbf{k}}}{\partial
k_x}[2\frac{f_{n\mathbf{k}}-f_{n'\mathbf{k}}}{(E_{n\mathbf{k}}-E_{n'\mathbf{k}})^{3}}\\
&-\frac{\partial f_{n'\mathbf{k}}/\partial
E_{n'\mathbf{k}}}{(E_{n\mathbf{k}}-E_{n'\mathbf{k}})^{2}}]+o(q_x^2).
\label{tor4}
\end{split}
\end{equation}
%Applying Eq. (\ref{tor3}) and Eq. (\ref{tor4}) to Eq. (\ref{tor2})
Assuming the Fermi energy is lager than the spin-orbit
splitting and using using
$(k^{+}_{F}-k^{-}_{F})=2m\gamma(\phi)/\hbar^2$ and
$\sigma^{\tau}_{xy}=-Im[\chi^{(1)}_y]$, we can write the torque conductivity
as
\begin{equation}
\begin{split}
\sigma^{\tau}_{xy}=&\frac{e\beta}{4\pi^2}\int^{2\pi}_{0}d\phi\frac{G(\lambda,\beta,\phi)}{\gamma(\phi)}\\
&-\frac{e\beta}{8\pi^2}\int^{2\pi}_{0}d\phi\frac{H(\lambda,\beta,\phi)\cos\phi}{[\gamma(\phi)]^2},
\label{Int1}
\end{split}
\end{equation}
where $G(\lambda,\beta,\phi)$ and $H(\lambda,\beta,\phi)$ are
given by
\begin{eqnarray*}
G(\lambda,\beta,\phi)&=&\frac{(\beta^2-\lambda^2)\sin^2\phi(\lambda
\sin(2\phi)-\beta)}{(\gamma(\phi))^3}\\
\\
H(\lambda,\beta,\phi)&=&\lambda \sin\phi+\beta
\cos\phi\\
&&+(\lambda^2-\beta^2)\frac{(\lambda \sin\phi-\beta
\cos\phi)\cos(2\phi)}{(\gamma(\phi))^2}\\
&&+(\lambda^2+\beta^2)\frac{(\lambda \cos\phi+\beta
\sin\phi)\sin(2\phi)}{(\gamma(\phi))^2}\\
&&-2\lambda\beta\frac{(\lambda \cos\phi+\beta
\sin\phi)}{(\gamma(\phi))^2}.
\end{eqnarray*}
Using
$[\gamma(\phi)]^2=(\lambda^2+\beta^2)-2\lambda\beta\sin(2\phi)$,
we find that
$\frac{G}{\gamma(\phi)}-\frac{H\cos\phi}{2\gamma(\phi)^2}$ can be
written as:
\begin{equation}
[\frac{G(\lambda,\beta,\phi)}{\gamma(\phi)}-\frac{1}{2}\frac{H(\lambda,\beta,\phi)}{\gamma(\phi)^2}]=(\beta^2-\lambda^2)\frac{2\lambda\sin\phi\cos\phi-\beta}{\gamma(\phi)^4}
\label{Int2}
\end{equation}
>From the above formula, we can easily check that Eq. (\ref{Int2})
the is zero when we set $\lambda^2=\beta^2$. From Eq.
(\ref{Int2}), it follows that the torque conductivity vanishes in
pure Rashba system (i.e. $\beta=0$) because the integral $\int
d\phi\sin\phi\cos\phi$ is zero. Our next step is to work out the
integrals by using Residue method~\cite{arf95}. The crucial
integrals are the followings ($|\lambda|
> |\beta|$):
\begin{equation}
    \begin{split}
&\int^{2\pi}_{0}d\phi\frac{1}{[\gamma(\phi)]^4}=\frac{2\pi(\lambda^2+\beta^2)}{(\lambda^2-\beta^2)^3}\\
&\int^{2\pi}_{0}d\phi\frac{\sin\phi \cos\phi}{[\gamma(\phi)]^4}=\frac{2\pi\lambda\beta}{(\lambda^2-\beta^2)^3}\\
\label{Int3}
\end{split}
\end{equation}
For $|\beta|>|\lambda|$, we can exchange the Rashba and
Dresselhaus couplings $\lambda\longleftrightarrow\beta$ in the
integrals. Substituting Eq. (\ref{Int2}) and Eq. (\ref{Int3}) into
Eq. (\ref{Int1}), we obtain the torque conductivity:
\begin{equation}
\sigma^{\tau}_{xy}=-\frac{e}{2\pi}\frac{\beta^2}{|\lambda^2-\beta^2|}.
\end{equation}
Since the commutation relation $[s_z+L_z,\frac{p^2}{2m}+H_R]=0$,
the Rashba coupling disappears in numerator of
$\sigma^{\tau}_{xy}$. For
$\tau^R_s\equiv\frac{1}{i\hbar}[s_z,H_R(\textbf{k})]=\lambda(\sigma_xk_x+\sigma_yk_y)$,
we can use the same steps as the calculation of
$\sigma^{\tau}_{xy}$. The new $H_s(\lambda,\beta,\phi)$ and
$G_s(\lambda,\beta,\phi)$ can be obtained by changing the
$\lambda$ and $\beta$ in $H(\lambda,\beta,\phi)$ and
$G(\lambda,\beta,\phi)$. We have
\begin{equation}
\begin{split}
\sigma^{\tau^R_s}_{xy}=&-\sigma^{\tau^R_{o}}_{xy}\\
=&-\frac{e\lambda}{8\pi^2}\int^{2\pi}_{0}d\phi\frac{G_s(\lambda,\beta,\phi)}{\gamma(\phi)}\\
&+\frac{e\lambda}{16\pi^2}\int^{2\pi}_{0}d\phi\frac{H_s(\lambda,\beta,\phi)\cos\phi}{[\gamma(\phi)]^2}\\
=&-\frac{e\lambda}{8\pi^2}\int^{2\pi}_{0}d\phi\frac{G(\lambda\leftrightarrow\beta,\phi)}{\gamma(\phi)}\\
&+\frac{e\lambda}{16\pi^2}\int^{2\pi}_{0}d\phi\frac{H(\lambda\leftrightarrow\beta,\phi)\cos\phi}{[\gamma(\phi)]^2}\\
=&\frac{e}{4\pi}\frac{\lambda^2}{|\lambda^2-\beta^2|}.\\
\end{split}
\end{equation}
Therefore, the spin torque Hall conductivity is
\begin{equation}
  \begin{split}
\sigma^{\tau_s}_{xy}&=(-\frac{e}{4\pi}\frac{\beta^2}{|\lambda^2-\beta^2|}+\frac{e}{4\pi}\frac{\lambda^2}{|\lambda^2-\beta^2|})\\
&=\frac{e}{4\pi}\frac{\lambda^2-\beta^2}{|\lambda^2-\beta^2|}=sign(\lambda^2-\beta^2)\frac{e}{4\pi}\\
   \end{split}
\end{equation}
and the OAM torque Hall conductivity is
\begin{equation}
    \begin{split}
\sigma^{\tau_{o}}_{xy}&=(-\frac{e}{4\pi}\frac{\beta^2}{|\lambda^2-\beta^2|}-\frac{e}{4\pi}\frac{\lambda^2}{|\lambda^2-\beta^2|})\\
&=-\frac{e}{4\pi}\frac{\lambda^2+\beta^2}{|\lambda^2-\beta^2|}.
    \end{split}
\end{equation}
For the pure Rashba system, we can take $\beta=0$ in the
$\lambda^2>\beta^2$ condition. For the pure Dresselhaus system, we can
take $\lambda=0$ in the $\lambda^2<\beta^2$ condition.

\section{Gauge invariant position operator}
In this appendix, we shall consider a method to obtain the gauge
invariant position operator and resolve the problem that the
expectation value of the OAM depends on the
choice of the eigenstates in the absence of applied electric field. In
crystalline environment, the position operator cannot be simply
set to
$\mathbf{x}=i\frac{\partial}{\partial\mathbf{k}}$~\cite{BLO62}. We
find that the OAM will depend on the choice
of the eigenstates when we use the definition:
$\mathbf{x}=i\frac{\partial}{\partial\mathbf{k}}$. For example,
the following states can also be chosen as the eigenstates
\begin{equation}
|n\mathbf{k};M\rangle=\frac{e^{iM(\mathbf{k})}}{\sqrt{2}}\begin{pmatrix}
1
\\ ine^{i\theta(\mathbf{k})}
\label{basisM}
\end{pmatrix}
\end{equation}
where $M(\mathbf{k})$ is the phase factor and $n=\pm 1$. It
can be proved that the eigenstates of Eq. (\ref{basisM}) satisfy the relation
$H_0=\sum_{n}E_{n\mathbf{k}}|n\mathbf{k};M\rangle\langle
n\mathbf{k};M|$ where the eigenenergy $E_{n\mathbf{k}}$ is defined
in Eq. (\ref{eigenenergies_RD}). Using the eigenstates of Eq. (\ref{basisM}),
the matrix element $\langle
n\mathbf{k};M|L_z|n\mathbf{k};M\rangle$ is zero if
$M=-\theta/2$ but nonzero if $M=-\theta$. Furthermore, if the eigenstates
of Eq. (\ref{eigenstate RD2}) are chosen, the signs of
$\langle\widetilde{+\mathbf{k}}|L_z|\widetilde{+\mathbf{k}}\rangle$
and
$\langle\widetilde{-\mathbf{k}}|L_z|\widetilde{-\mathbf{k}}\rangle$
will be different. Let us now consider the phase transformation of
$|\overline{n\mathbf{k}}\rangle=e^{i\Lambda_{n}(\mathbf{k})}|n\mathbf{k}\rangle$
where the phase $\Lambda_n(\mathbf{k})$ depends on band index
$n$. We note that the eigenstates of Eq. (\ref{eigenstate_RD})
will be transformed to that of Eq. (\ref{eigenstate RD2}) by the appropriate
choice of the phase $\Lambda_{n}$. Actually, we have
$|\widetilde{+\mathbf{k}}\rangle=(-i)|+\mathbf{k}\rangle$ and
$|\widetilde{-\mathbf{k}}\rangle=e^{i\theta}|-\mathbf{k}\rangle$.
Let us also introduce the operator $\mathbf{X}'\equiv
\mathbf{x}+\mathbf{A}'$ where
$\mathbf{x}=i\frac{\partial}{\partial\mathbf{k}}$ and
$\mathbf{A}'$ is the vector potential. Using the above
gauge transformation~\cite{boh03}, we obtain
\begin{equation}
\langle\overline{n\mathbf{k}}|\mathbf{X}'|\overline{n\mathbf{k}}\rangle=\langle
n\mathbf{k}|\mathbf{x}|n\mathbf{k}\rangle+\mathbf{A}'-\frac{\partial\Lambda_n}{\partial\mathbf{k}}.
\label{GT1}
\end{equation}
Consider the specific transformation:
\begin{equation}
\mathbf{A}'=\mathbf{A}+\frac{\partial\Lambda_n}{\partial\mathbf{k}}.
\label{GT2}
\end{equation}
Eq. (\ref{GT1}) becomes the gauge invariant form~\cite{Sak94}:
\begin{equation}
\langle\overline{n\mathbf{k}}|\mathbf{X}'|\overline{n\mathbf{k}}\rangle=\langle
n\mathbf{k}|\mathbf{X}|n\mathbf{k}\rangle,
\label{GT3}
\end{equation}
where the operator $\mathbf{X}$ is defined as
$\mathbf{X}=i\frac{\partial}{\partial\mathbf{k}}+\mathbf{A}$. We
find that $\mathbf{A}'$ satisfies the transformation (\ref{GT2}) if
$\mathbf{A}'$ is defined as
\begin{equation}
\mathbf{A}_{n}'\equiv\langle
\overline{n\mathbf{k}}|(-i)\frac{\partial}{\partial
    \mathbf{k}}|\overline{n\mathbf{k}}\rangle
\label{GT4}
\end{equation}
which is the Berry vector potential when the eigenstates
$|\overline{n\mathbf{k}}\rangle$ are used. The vector potential
$\mathbf{A}$ in the eigenstates $|n\mathbf{k}\rangle$ can be
defined as $\mathbf{A}_{n}=\langle
n\mathbf{k}|(-i)\frac{\partial}{\partial\mathbf{k}}|n\mathbf{k}\rangle$
which generally depends on band index $n$. Therefore, the gauge
invariant position operator $\mathbf{X}$ will depend on band index
$n$:
\begin{equation}
\mathbf{X}_{\pm}=i\frac{\partial}{\partial\mathbf{k}}+\mathbf{A}_{\pm}.
\end{equation}
We find that the expectation value $\langle
n\mathbf{k}|\mathbf{X}_{n}|n\mathbf{k}\rangle$ vanishes: $\langle
n\mathbf{k}|\mathbf{X}_{n}|n\mathbf{k}\rangle=0$. It can also be
shown that $\langle n'\neq
n\mathbf{k}|\textbf{X}_{n}|n\mathbf{k}\rangle=\langle n'\neq
n\mathbf{k}|i\frac{\partial}{\partial
\mathbf{k}}|n\mathbf{k}\rangle$ by the use of $\langle
n'\mathbf{k}|n\mathbf{k}\rangle=\delta_{n'n}$ where $\delta_{n'n}$
is equal to unity for $n=n'$ and zero for $n\neq n'$. In terms of
the gauge invariant position operator, the OAM
operator $(\mathbf{x}\times\mathbf{k})_z$ is replaced by
$(\mathbf{X}_{n}\times\mathbf{k})_z$.
% which depends on band index $n$.
As a result, the expectation value of the OAM operator
$\langle\mathbf{L}\rangle_0=\hbar\langle
n\mathbf{k}|\mathbf{X}_{n}|n\mathbf{k}\rangle\times\mathbf{k}$ is
zero for all the eigenstates in the absence of the applied electric field.
Interestingly, note that the OAM operator does not always
depend on band index $n$. For example, if the eigenstates of Eq.
(\ref{eigenstate_RD}) are used, the Berry vector potential is independent
of $n$, and so is the gauge invariant operator
$\mathbf{X}_n$.
%Conversely, the
%operator $\mathbf{X}_{n}$ will depend on $n$ when we use the basis
%(\ref{eigenstate RD2}).

Let us now show that the calculated OAM Hall conductivities
remain unchanged no matter whether the conventional $\mathbf{x}$ or
gauge invariant $\mathbf{X}_n$ position operator is used.
%consider the velocity, OAM torque, and OAM current operators
%that appear in Eq. (\ref{equ:kubo_1}) and Eq. (\ref{equ:kubo_2}).
%We will obtain the same conclusions no matter the $\mathbf{X}_{-}$
%or $\mathbf{X}_{+}$ is used.
First of all, the gauge invariant velocity operator
$\frac{1}{i\hbar}[\mathbf{X}_{\pm},H_0(\mathbf{k})]$ is the same
as the original one
$\frac{1}{i\hbar}[\mathbf{x},H_0(\mathbf{k})]$ because the vector
potential $\mathbf{A}_{\pm}$ commutes with $H_0(\mathbf{k})$,
namely, $[\mathbf{A}_{\pm},H_0(\mathbf{k})]=0$.
%where $H_0$ is defined by Eq. (\ref{hamiltonian}).
The gauge invariant OAM torque operator
$\frac{1}{i\hbar}[(\mathbf{X}_{\pm}\times\mathbf{k})_z,H_0]$
is also the same as the conventional one
$\frac{1}{i\hbar}[(\mathbf{x}\times\mathbf{k})_z,H_0]$ because
$(\mathbf{A}_{\pm}\times\mathbf{k})_z$ commutes with
$H_{0}(\mathbf{k})$. Nevertheless, the OAM current operator
$\frac{1}{2}\{v_x,L_z\}$ has an extra term of
$(\mathbf{A}_{\pm}\times\mathbf{k})_zv_x$ when the gauge invariant
position operator
$\mathbf{X}_{\pm}$ is used. For the eigenstates of Eq.
(\ref{basisM}), we find that $\langle
n\mathbf{k};M|v_x|-n\mathbf{k};M\rangle\langle
-n\mathbf{k};M|v_y|n\mathbf{k};M\rangle=-[(\lambda^2+\beta^2)\sin\theta\cos\theta+\lambda\beta]/\hbar^2$
and $\mathbf{A}_{\pm}=\frac{1}{2}(2\frac{\partial
M}{\partial\mathbf{k}}+\frac{\partial\theta}{\partial\mathbf{k}})$
are all purely real \cite{Bvec}. On the other hand, Eq.
(\ref{equ:kubo_1}) needs only the imaginary part of the matrix elements
$(\mathbf{A}_{\pm}\times\mathbf{k})_{z}\langle
n\mathbf{k};M|v_x|-n\mathbf{k};M\rangle\langle
-n\mathbf{k};M|v_y|n\mathbf{k};M\rangle$. This shows that the
extra term $(\mathbf{A}_{\pm}\times\mathbf{k})_zv_x$ does not
contribute to the Kubo formula. In general, using
$|\overline{n\mathbf{k}}\rangle=e^{i\Lambda_{n}}|n\mathbf{k}\rangle$,
we have $\langle
\overline{n\mathbf{k}}|v_x|\overline{n'\mathbf{k}}\rangle\langle
\overline{n'\mathbf{k}}|v_y|\overline{n\mathbf{k}}\rangle=\langle
n\mathbf{k}|v_x|n'\mathbf{k}\rangle\langle
n'\mathbf{k}|v_y|n\mathbf{k}\rangle$. With the eigenstates of Eq.
(\ref{eigenstate RD2}), we obtain that $\langle
\widetilde{n\mathbf{k}}|v_x|\widetilde{-n\mathbf{k}}\rangle\langle
\widetilde{-n\mathbf{k}}|v_y|\widetilde{n\mathbf{k}}\rangle=-[(\lambda^2+\beta^2)\sin\theta\cos\theta+\lambda\beta]/\hbar^2$
and
$\mathbf{A}_{\pm}=\mp\frac{1}{2}\frac{\partial\theta}{\partial\mathbf{k}}$,
which are all purely real. Therefore, we conclude that the OAM Hall
conductivities $\sigma^{o_0}_{xy}$ and $\sigma^{\tau_o}_{xy}$
remain the same even if we replace
$\mathbf{x}$ with $\mathbf{X}_{+}$ or $\mathbf{X}_{-}$.

%\bibliography{ref}

\begin{references}
\bibitem{cond} We evaluate $\sigma^{s_0}_{yx}$ and $\sigma^{o_0}_{yx}$ directly,
and find the relations of
$\sigma^{s_0}_{xy}=-\sigma^{s_0}_{yx}$ and
$\sigma^{o_0}_{xy}=-\sigma^{o_0}_{yx}$.
\bibitem{Bvec} In general, it can be proved that $\langle
n\mathbf{k}|\frac{\partial}{\partial\mathbf{k}}|n\mathbf{k}\rangle$
is purely imaginary by means of $\langle
n\mathbf{k}|n\mathbf{k}\rangle=1$ (see Ref. \onlinecite{boh03}).
In our definition of $\mathbf{A}_{n}=\langle
n\mathbf{k}|(-i)\frac{\partial}{\partial\mathbf{k}}|n\mathbf{k}\rangle$,
however, $\mathbf{A}_{n}$ is purely real.
\end{references}

\end{document}